\documentclass[a4paper,11pt]{JHEP}
\usepackage[T1]{fontenc}		
\usepackage{ae,aecompl,aeguill}
\usepackage[ansinew]{inputenc}
\usepackage{amsmath}
\usepackage{amssymb}
\usepackage{slashed}
\usepackage{graphicx}
\usepackage{subfigure}

\newcommand{\be}{\begin{equation}}
\newcommand{\ee}{\end{equation}}
\newcommand{\beqa}{\begin{eqnarray}}
\newcommand{\eeqa}{\end{eqnarray}}

\newcommand\di{\mathrm{d}}

\newcommand\ncdm{{\tt ncdm} }
\newcommand{\para}[1]{\left( #1 \right)}
\newcommand{\bpara}[1]{\left[ #1 \right]}
\newcommand{\qmin}{q_\text{min}}
\newcommand{\qmax}{q_\text{max}}
\newcommand{\eV}{\text{eV}}
\newcommand\CLASS{{\tt CLASS}}
\newcommand\CAMB{{\tt CAMB}}
\newcommand\CMBFAST{{\tt CMBFAST}}
\newcommand\CMBEASY{{\tt CMBEASY}}

\preprint{CERN-PH-TH/2011-084, LAPTH-012/11} \title{The Cosmic Linear
Anisotropy Solving System\\ (CLASS) IV: Efficient implementation of
non-cold relics} \author{Julien Lesgourgues$^{a,b,c}$, Thomas
Tram$^{d,b}$\vspace{.2cm}\\ {$^a$}Institut de Th\'eorie des
Ph\'enom\`enes Physiques,\\ \'Ecole Polytechnique F\'ed\'erale de
Lausanne,\\ CH-1015, Lausanne, Switzerland.\vspace{.2cm}\\ {$^b$}
CERN, Theory Division,\\ CH-1211 Geneva 23,
Switzerland.\vspace{.2cm}\\ {$^c$} LAPTh (CNRS - Universit\'e de
Savoie), BP 110,\\ F-74941 Annecy-le-Vieux Cedex,
France.\vspace{.2cm}\\ {$^d$} Department of Physics and Astronomy,\\
University of Aarhus,\\ DK-8000 Aarhus C, Denmark.}

\abstract{We present a new flexible, fast and accurate way to
implement massive neutrinos, warm dark matter and any other non-cold
dark matter relics in Boltzmann codes.  For whatever analytical or
numerical form of the phase-space distribution function, the optimal
sampling in momentum space compatible with a given level of accuracy
is automatically found by comparing quadrature methods.  The
perturbation integration is made even faster by switching to an
approximate viscous fluid description inside the Hubble radius, which
differs from previous approximations discussed in the literature. When
adding one massive neutrino to the minimal cosmological model,
\CLASS{} becomes just 1.5 times slower, instead of about 5 times in
other codes (for fixed accuracy requirements). We illustrate the
flexibility of our approach by considering a few examples of standard
or non-standard neutrinos, as well as warm dark matter models.}

\begin{document} 

\section{Introduction}
The inclusion of massive, non-cold relics in a Boltzmann code is
complicated by the fact that it is necessary to evolve the
perturbation of the distribution function on a momentum grid. A grid
size of $N$ points together with $L$ terms in the expansion of the
perturbation leads to $N\cdot L$ added equations to the system. In
public Boltzmann codes like \CMBFAST{}~\cite{Seljak:1996is}, \CAMB{}~\cite{Lewis:1999bs}  and \CMBEASY~\cite{Doran:2003sy}, distributions are
sampled evenly with fixed step size and maximum
momentum, adapted to the case of a Fermi-Dirac shaped distribution
function $f(p)$. Moreover, the analytic expression for $f(p)$ is
hard-coded in many places in those codes, and implicitly assumed
e.g. in the mass to density relation, so that exploring other models
like neutrinos with chemical potentials and flavour oscillations, neutrinos
with non-thermal corrections, extra sterile neutrinos or any kind of
warm dark matter candidate requires non-trivial changes to these
codes.

Throughout this paper, when discussing \CAMB{} or \CLASS{}, we refer to the versions available at the time of preparing this manuscript, i.e. the January 2011 version of \CAMB{} and {\tt v1.1} of \CLASS{}. Note that a handful of improvements on the massive neutrino implementation in \CAMB{} was added to the July 2011 version, some of which were inspired by the current work.

We present here the way in which generic Non-Cold Dark Matter (NCDM)
relics are implemented in the new Boltzmann code
\CLASS{}\footnote{available at {\tt http://class-code.net}. This paper
is based on version {\tt v1.1} of the code.} (Cosmic Linear Anisotropy
Solving System), already presented in a series of companion
papers~\cite{class_gen,class_approx,class_comp}. In order to ensure a
complete flexibility, \CLASS{} assumes an arbitrary number of NCDM
species, each with an arbitrary distribution function $f_i(p)$. For
each species, this function can be passed by the user under some
(arbitrarily complicated) analytic form in a unique place in the code,
or in a file in the case of non-trivial scenarios that requires a
numerical simulation of the freeze-out process. All other steps
(finding a mass-density relation, optimising the momentum sampling and
computing the derivative of $f_i(p)$) are done automatically in order
to ensure maximum flexibility.

In Sec.~\ref{sampling} of the, we present an automatic quadrature method
comparison scheme which allows \CLASS{} to find an optimal momentum
sampling, given $f_i(p)$ and some accuracy requirement, and in 
Sec.~\ref{ncdmfa}, we devise a new approximation scheme allowing us to
drastically reduce the computational time for wavelengths inside the
Hubble radius. Finally, in Sec.~\ref{neutrinos}~and~\ref{exemples}, we illustrate these
methods with several examples based on standard and non-standard massive
neutrinos, and different types of warm dark matter candidates.

\section{Massive neutrino perturbations}

The formalism describing the evolution of any NCDM species is given by
the massive neutrino equations of Ma \& Bertschinger \cite{Ma:1995ey}.
We will follow the notations from this paper closely, with the exceptions
\begin{align}
q &\equiv \frac{q_\text{\tt MB}}{T_{\ncdm,0}}~, \qquad 
\epsilon \equiv \frac{\epsilon_\text{\tt MB}}{T_{\ncdm,0}} = \para{q^2 + a^2\frac{m^2}{T_{\ncdm,0}^2}}^{\frac{1}{2}},
\end{align}
where $a$ is the scale factor, and $m$ and $T_{\ncdm,0}$ is the mass and the current temperature of the non-cold relic, in the case of a thermal relic. If the relic is non-thermal, $T_{\ncdm,0}$ is just a scale of the typical physical momentum of the particles today. Note that the perturbation equations Eq.~\eqref{eq:boltzmann} are still the same as in~\cite{Ma:1995ey}, since they depend only on the ratio $q/\epsilon$ which is not affected by this rescaling.

\subsection{Perturbations on a grid}

We are not interested in the individual momentum components of the
perturbation, $\Psi_l$, but only in the perturbed energy density,
pressure, energy flux and shear stress of each NCDM species, which are
integrals over $\Psi_l$~\cite{Ma:1995ey}:
\begin{subequations}
\label{eq:perturbations}
\begin{align}
\delta\rho_\ncdm &= 4\pi \para{\frac{T_{\ncdm,0}}{a}}^4 \int_0^\infty f_0(q)\di q q^2\epsilon \Psi_0, \label{eq:deltancdm}\\
\delta p_\ncdm &= \frac{4\pi}{3} \para{\frac{T_{\ncdm,0}}{a}}^4 \int_0^\infty f_0(q)\di q \frac{q^4}{\epsilon} \Psi_0, \label{eq:deltap} \\
\left( \bar{\rho}_\ncdm + \bar{p}_\ncdm \right) \theta_\ncdm &=
	4\pi k \para{\frac{T_{\ncdm,0}}{a}}^4 \int_0^\infty f_0(q)\di q q^3 \Psi_1, \label{eq:thetancdm}\\
\left( \bar{\rho}_\ncdm + \bar{p}_\ncdm \right) \sigma_\ncdm &=
	\frac{8\pi}{3} \para{\frac{T_{\ncdm,0}}{a}}^4 \int_0^\infty f_0(q)\di q \frac{q^4}{\epsilon} \Psi_2.\label{eq:sigmancdm}
\end{align}
\end{subequations}
In the rest of the article, we will omit all $\ncdm$ subscripts, and
dots will denote derivatives with respect to conformal time, $\tau$. In Eq.~\eqref{eq:perturbations} and elsewhere, $f_0(q)$ is the unperturbed phase-space-distribution of the non-cold species.

Note that $\Psi_0$ and $\Psi_1$ are gauge-dependent quantities, while
higher momenta are not. The gauge transformation can be derived from the corresponding gauge transformation of the integrated quantities. The relation between $\Psi_1$ in the conformal Newtonian gauge and in the synchronous
one reads:
\begin{equation}
\Psi_{1,\text{Con.}} = \Psi_{1,\text{Syn.}} - \frac{1}{3} \alpha k
\frac{\epsilon}{q}\frac{d \ln f_0}{d \ln q}~,
\end{equation}
with $\alpha \equiv (\dot{h} + 6 \dot{\eta})/(2k^2)$, where $h$ and
$\eta$ are the usual scalar metric perturbations in the synchronous gauge.  In
the rest of this paper, we will work exclusively in the synchronous gauge.  The
evolution of the $\Psi_l$'s are governed by the Boltzmann equation as
described in~\cite{Ma:1995ey}, and leads to the following system of
equations:
\begin{subequations} \label{eq:boltzmann}
\begin{align}
\dot{\Psi}_0 &= -\frac{qk}{\epsilon} \Psi_1 + \frac{\dot{h}}{6}\frac{\di \ln f_0}{\di \ln q}, \\
\dot{\Psi}_1 &= \frac{qk}{3\epsilon} \para{\Psi_0 -2 \Psi_2}, \\
\dot{\Psi}_2 &= \frac{qk}{5\epsilon} \para{2\Psi_1-3\Psi_3} -\para{\frac{\dot{h}}{15} + \frac{2\dot{\eta}}{5} } \frac{\di \ln f_0}{\di \ln q}, \label{eq:psi2dot}\\
\dot{\Psi}_{l\geq 3} &= \frac{qk}{\para{2l+1}\epsilon} \para{ l\Psi_{l-1} -\para{l+1} \Psi_{l+1}}.
\end{align}
\end{subequations}
We can write the homogeneous part of this set of equations as
\begin{align}
\mathbf{\dot{\Psi}} &= \frac{qk}{\epsilon}A \mathbf{\Psi} \equiv \alpha\para{\tau} A \mathbf{\Psi},
\end{align}
where $A$ is given by
\begin{align}
A &= 
\begin{bmatrix}
  &  -1 &  & & &\\
  \frac{1}{3} & & -\frac{2}{3} & & &\\
  & \ddots & & \ddots & &\\
  & & \frac{l}{2l+1} & & -\frac{l+1}{2l+1} &\\
  & & & \ddots & & \ddots \\
  & & & & \ddots & &
\end{bmatrix}
\end{align}
The solution can be written in terms of the matrix exponential,
\begin{align}
\mathbf{\Psi}\para{\tau} &= e^{\int_{\tau_i}^\tau \di \tau '\alpha\para{\tau '}A}\mathbf{\Psi}\para{\tau_i} \\
&= U e^{\int_{\tau_i}^\tau \di \tau '\alpha\para{\tau '}D} U^{-1} \mathbf{\Psi}\para{\tau_i},
\end{align}
where $A$ has been diagonalised such that $A = UDU^{-1}$ and $D$ is a diagonal matrix of eigenvalues of A. The largest eigenvalue of $A$ (using the complex norm) goes toward $\pm i$ for $l_\text{max} \rightarrow \infty$, so the time-dependent phase corresponding to the largest frequency oscillation in the system, is given by
\begin{align}
\phi_{\omega_\text{max}}\para{\tau} \simeq k \int_{\tau_i}^\tau \di \tau '\para{1+\frac{M^2}{q^2}a\para{\tau '}^2}^{-\frac{1}{2}}. \label{eq:wmax}
\end{align}

\subsection{Quadrature strategy}
There is no coupling between the momentum bins, so our only concern is
to perform the indefinite integrals numerically with sufficient accuracy while using the
fewest possible points. We are interested in the integrals in Eq.~\eqref{eq:perturbations}, which are all on the form
\begin{align}
\mathcal{I} &= \int_0^\infty \di q f_0\para{q} g\para{q},
\end{align}
where $f_0(q)$ is the phase space distribution and $g(q)$ is some function of $q$. We will assume that $g(q)$ is reasonably well described by a polynomial in $q$, which we checked explicitly for the functions in Eq.~\eqref{eq:perturbations}. Under this assumption, we can determine the accuracy of any quadrature rule on $\mathcal{I}$ by performing the integral
\begin{align}
\mathcal{J} &= \int_0^\infty \di q f_0\para{q} t\para{q},
\end{align}
where $t(q)$ is a test function. Given a set of different quadrature rules for performing the integral $\mathcal{I}$, the idea is to choose the rule which can compute $\mathcal{J}$ to the required accuracy {\tt tol\_ncdm} using the fewest possible points. The details of our quadrature methods can be found in Sec.~\ref{sampling} of the appendix.

Note that higher accuracy is needed for integrating background
quantities (density, pressure, etc.) than perturbed quantities (the
$\Psi_l$'s). On the other hand, the code spends a negligible time in
the computation of the former, while reducing the number of sampling
points for perturbations is crucial for reducing the total computing
time.  Hence, \CLASS{} calls the quadrature optimisation algorithm twice
for each NCDM species, with two different accuracy parameters.  The
background tolerance is set to a smaller value leading to a finer
sampling.

The process of finding the optimal roots and weights is quite involved, but it requires a negligible computing time in
\CLASS{}. What really matters is to reduce the number of discrete momenta
in the perturbation equations, and this is indeed accomplished thanks to
the previous steps (as we shall see in Sec.~\ref{exemples}).

\section{Sub-Hubble Approximation \label{ncdmfa}}
\subsection{Fluid approximation}

Various kinds of approximations for massive neutrino perturbations
have been discussed in the past \cite{Hu:1998kj,Lewis:2002nc,Shoji:2010hm}. The
approximation discussed here is different and consists in an
extension of the Ultra-relativistic Fluid Approximation presented
in~\cite{class_approx}, applying only to the regime in which a
given mode has entered the Hubble radius.  The idea is that after
Hubble crossing, there is an effective decoupling between high
multipoles (for which power transfers from smaller $l$'s to higher
$l$'s, according to the free-streaming limit) and low multipoles
(just sourced by metric perturbation). Hence, when $k\tau$ exceeds
some threshold, we can reduce the maximum number of multipoles from
some high $l_{\rm max}$ down to $l_{\rm max}=2$. We showed in~\cite{class_approx} that this Ultra-relativistic Fluid Approximation
(UFA) allows simultaneously to save computing time (by reducing the number of
equations) and to increase precision (by avoiding artificial
reflection of power at some large cut-off value $l_{\rm max}$).

In the case of massive neutrinos, we expect the same arguments to hold
in the relativistic regime, while in the non-relativistic limit all
multipoles with $l>1$ decay and the species behave more and more like
a pressureless fluid. Hence, some kind of fluid approximation is
expected to give good results in all cases.

We write the continuity equation and the Euler equation
in the usual way. In the synchronous gauge we have
\begin{subequations}
\begin{align}
\dot{\delta} &= -\para{1+w}\para{\theta + \frac{\dot{h}}{2}} -3\frac{\dot{a}}{a}\para{c_\text{Syn.}^2-w}\delta, \\
\dot{\theta} &= -\frac{\dot{a}}{a}\para{1-3c_g^2}\theta + \frac{c_\text{Syn.}^2}{1+w}k^2\delta-k^2\sigma. 
\end{align}
\end{subequations}
Here, $c_g^2\equiv \frac{\dot{p}}{\dot{\rho}}$ is the adiabatic sound speed, and $c_\text{Syn.}^2 \equiv
\frac{\delta p}{\delta \rho}$ is the effective sound speed squared in the
synchronous gauge. The latter can be related to the physical sound speed
defined in the gauge comoving with the fluid, that we denote
$c_\text{eff}$. The above equations can then be written as:
\begin{subequations}
\begin{align}
\dot{\delta} &= -\para{1+w}\para{\theta + \frac{\dot{h}}{2}} -3\frac{\dot{a}}{a}\para{c_\text{eff}^2-w}\delta
+9\para{\frac{\dot{a}}{a}}^2 (1+w) \para{c_\text{eff}^2-c_g^2}\frac{\theta}{k^2}, \label{eq:continuity} \\
\dot{\theta} &= -\frac{\dot{a}}{a}\para{1-3c_\text{eff}^2}\theta + \frac{c_\text{eff}^2}{1+w}k^2\delta-k^2\sigma. \label{eq:Euler} 
\end{align}
\end{subequations}
Later on, we will close the system by an evolution equation for the shear $\sigma$, but first we will discuss how to calculate the adiabatic sound speed and how to approximate the effective sound speed $c_\text{eff}^2$.

\subsection{Sound speeds}
The adiabatic sound speed can be expressed as
\begin{align}
c_g^2 &= \frac{\dot{p}}{\dot{\rho}} = w \frac{\dot{p}}{p}\para{\frac{\dot{\rho}}{\rho}}^{-1} =- w \frac{\dot{p}}{p}\para{\frac{\dot{a}}{a}}^{-1} \frac{1}{3\para{1+w}} \nonumber \\
	&=  \frac{w}{3\para{1+w}} \para{5-\frac{\mathfrak{p}}{p}},
\label{eq:ca2}
\end{align}
where the quantity  $\mathfrak{p}$ (called the pseudo-pressure inside \CLASS{}) is a higher moment pressure defined by
\begin{align}
\mathfrak{p} &\equiv \frac{4\pi}{3} \para{\frac{T_{\ncdm,0}}{a}}^4 \int_0^\infty f_0(q)\di q \frac{q^6}{\epsilon^3}. \label{eq:pseudopressure}
\end{align}
With this formulation, we can compute the adiabatic sound speed in a
stable and accurate way, without needing to evaluate the
time-derivative of the background pressure $\dot{p}$. 
When the \ncdm{} species is no longer relativistic, its pressure perturbation $\delta p$ defined in Eq.~\eqref{eq:deltap} is an independent quantity. Since we do not have an evolution equation for $\delta p$, we approximate $c_\text{eff}^2$ by $c_g^2$. This approximation is sometimes as much as a factor 2 wrong as shown on~Fig.~\ref{fig:ceff2}.

We tried to use an ad hoc fitting formula for $c_\text{eff}^2$ to quantify the impact of this approximation, and we found that although there was an improvement at the level of perturbations, the overall error was still at the same level. We conclude, that the error we make from this approximation is at least not dominating the total error of our approximation scheme.
\FIGURE{
%
\includegraphics[width=0.48\columnwidth]{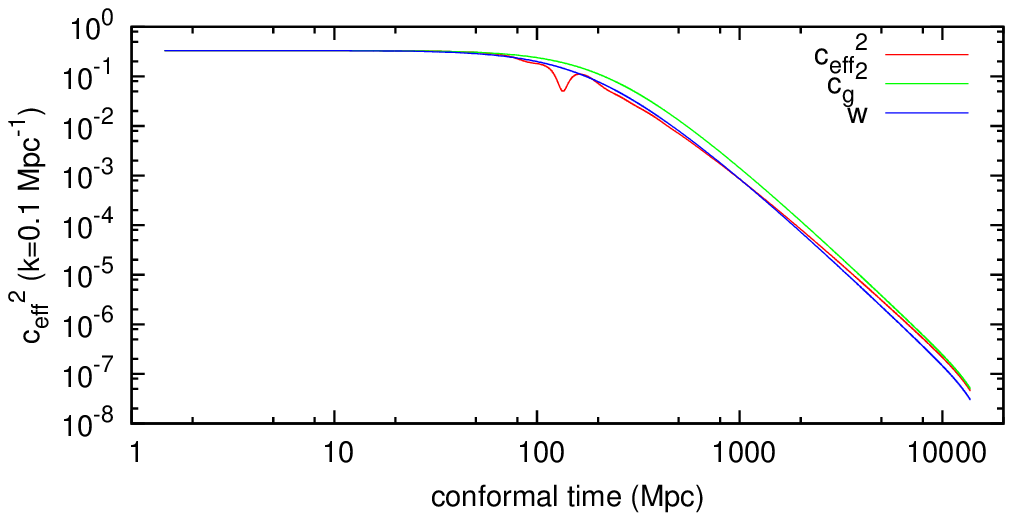}
\includegraphics[width=0.48\columnwidth]{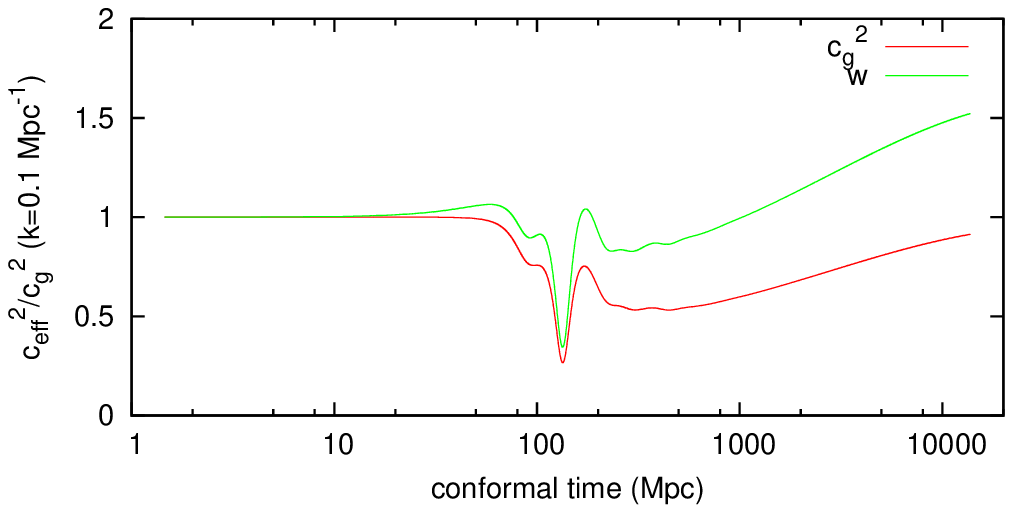}
\caption{\label{fig:ceff2} Effective sound speed squared
$c_\text{eff}^2$ for a mass of $m=2.0~\eV{}$. \emph{Left panel:} The
effective sound speed plotted together with the adiabatic sound speed
squared $c_g^2$ and the equation of state parameter $w$. In the relativistic
and the non-relativistic limit we have $c_\text{eff}^2 = c_g^2$ as
expected, but the behaviour of $c_\text{eff}^2$ in between the two
limits are non-trivial. \emph{Right panel:} The ratios
$c_\text{eff}^2/c_g^2$ and $c_\text{eff}^2/w$. One can see that
$c_g^2$ is a better approximation to $c_\text{eff}^2$ than $w$, but
neither catches the full evolution.}
}
\subsection{Evolution equation for the shear}
Given an ansatz for $\Psi_3$, we can derive a formally correct
  evolution equation for the shear. We follow Ma and Bertschinger, and
  close the system using their suggested recurrence relation for
  massive neutrinos\footnote{The recurrence relation in the massless
  limit is better motivated theoretically, since $\Psi_l \propto
  j_l\para{k\tau}$ when metric perturbations vanish or satisfy a
  simple constraint (namely, $\dot{\phi} + \dot{\psi} = 0$ in the
  Newtonian gauge). In the massive case, the formal solution involves
  more complicated oscillating functions with arguments going from $\sim k\tau$ in
  the massless limit to $\sim \para{k\tau}^{-1}$ in the massive limit,
  as can be checked from eq.~\eqref{eq:wmax}.}. The truncation law
  presented in Ma and Bertschinger is valid for $l_\text{max}>3$: in
  this case, all quantities are gauge-invariant. When writing the same
  ansatz for $l_\text{max}=3$, we have to face the issue of the gauge
  dependence of $\Psi_1$. Asssuming that the truncation law holds for
  gauge-invariant quantities, one obtains in the synchronous gauge:
\begin{equation}
\Psi_{3} \approx \frac{5 \epsilon}{qk\tau} \Psi_{2} - \para{\Psi_{1}
- \frac{1}{3} \alpha k \frac{\epsilon}{q} \frac{d \ln f_0}{d \ln q}}. 
\label{eq:recurrence}
\end{equation}
Throughout this subsection and the next one, one can recover Newtonian gauge equations by simply taking $\alpha=0$.
We now differentiate equation
\eqref{eq:sigmancdm}:
\begin{equation}
\dot{\sigma} + \frac{\dot{a}}{a}\para{1-3c_g^2} \sigma = \frac{1}{\rho
+ p} \frac{8\pi}{3}\para{\frac{T_{\ncdm,0}}{a}}^4 \int_0^\infty f_0(q)\di q q^4
\frac{\partial}{\partial\tau}\para{
\frac{\Psi_2}{\epsilon}}~. 
\end{equation} 
We can compute the
right-hand side using Eq.~\eqref{eq:psi2dot} and replace
$\Psi_3$ with its approximate expression from~\eqref{eq:recurrence}.
After carrying out integrals over momentum, one gets:
\begin{equation}
\dot{\sigma} =-3\para{\tau^{-1}+ \frac{\dot{a}}{a}\bpara{\frac{2}{3}-c_g^2-\frac{1}{3}\frac{\Sigma}{\sigma}}}\sigma + \frac{2}{3} \bpara{\Theta+ \alpha k^2 \frac{w}{1+w} \para{5 - \frac{\mathfrak{p}}{p}}}, \label{eq:shear_exact_con}
\end{equation}
where we have borrowed the notation
\begin{align}
\para{\rho + p} \Theta &=
	4\pi k \para{\frac{T_{\ncdm,0}}{a}}^4 \int_0^\infty f_0(q)\di q q^3 \frac{q^2}{\epsilon^2} \Psi_1, \label{eq:big_sigma} \\
\para{\rho + p} \Sigma &= \frac{8\pi}{3} \para{\frac{T_{\ncdm,0}}{a}}^4\int_0^\infty f_0(q)\di q \frac{q^4}{\epsilon} \frac{q^2}{\epsilon^2} \Psi_2, \label{eq:big_theta}
\end{align}
from~\cite{Shoji:2010hm}. 
From the definition it is clear that $\Theta \rightarrow \theta$ and
$\Sigma \rightarrow\sigma$ in the relativistic limit, and that
$\Theta$ and $\Sigma$ become suppressed in the non-relativistic
regime compared to $\theta$ and $\sigma$. Our differential equation
for $\sigma$ differs from its Newtonian gauge counterpart
in~\cite{Shoji:2010hm}, because we have used the recurrence relation
to truncate the hierarchy, while Shoji and Komatsu have used $\Psi_3 =
0$. The evolution equation for the shear can be further simplified by using 
Eq.~\eqref{eq:ca2}, leading to:
\begin{equation}
\dot{\sigma} =-3\para{\tau^{-1}+ \frac{\dot{a}}{a}\bpara{\frac{2}{3}-c_g^2-\frac{1}{3}\frac{\Sigma}{\sigma}}}\sigma + \frac{2}{3} \bpara{\Theta + 3 c_g^2 \alpha k^2}~.
\end{equation}

\subsection{Estimating quantities of higher velocity weight}

One way to close the system governing the fluid approximation is to
replace $\Theta$ and $\Sigma$ by the usual quantities $\theta$ and
$\sigma$ multiplied by functions depending only on background
quantities (in the same way that we already approximated $\delta p$ by
$c_g^2 \delta \rho$).  More explicitly, our aim is to write
an approximation of the type
$\Sigma = 3 w_\sigma \sigma$, where $w_\sigma$ could be any function
of time going from one third in the relativistic limit to zero in the
non-relativistic one. Since $\theta$ and $\Theta$ are not
gauge-independent, we should search for a similar approximation
holding on their gauge-independent counterpart. In the synchronous
gauge, such an approximation would read
\begin{equation}
\bpara{\Theta + 3 c_g^2 \alpha k^2 }
= 3 w_\theta \bpara{\theta + \alpha k^2}~.
\end{equation}
However, we will stick to the notations of \cite{Hu:1998kj}, who introduced a viscosity speed related to our $w_\theta$ through
\begin{equation}
c_\text{vis}^2 = \frac{3}{4} w_\theta (1 + w)~.
\end{equation}
With such assumptions, the approximate equation for the shear would read
\begin{equation}
\dot{\sigma} =-3\para{\frac{1}{\tau}+
\frac{\dot{a}}{a}\bpara{\frac{2}{3}-c_g^2-w_\sigma}}\sigma +
\frac{4}{3} \frac{c_\text{vis}^2}{1+w} \bpara{2 \theta + 2 \alpha
k^2}~. \label{eq:shear_exact_syn}
\end{equation}
Since the suppression factor $q^2/\epsilon^2$ which appears
in Eq.~(\ref{eq:big_sigma}, \ref{eq:big_theta}) compared
to Eq.~(\ref{eq:thetancdm}, \ref{eq:sigmancdm}) is also found in the
pressure integral compared to the energy density integral, we may
guess that the relative behaviour is similar, i.e. related by $w$. This leads to a guess $w_\sigma = w$ and $w_\theta
= w$ which implies $c_\text{vis}^2 =
\frac{3}{4}w\para{1+w}$. However, the same logic would imply $c_\text{eff}^2
= w$, which we have shown in Fig.~\ref{fig:ceff2} is not exactly true.

Let us investigate a bit how to approximate higher momenta quantities
like $\Theta$ and $\Sigma$. If we want to approximate $\Theta$ for
instance, we may assume some functional form of $\Psi_1(q)$ described
by a single (time dependent) parameter. We can make the ansatz
$\Psi_1(q/\epsilon) \approx a_{1n}(t) \para{\frac{q}{\epsilon}}^n$,
and then use $\theta$ to determine the parameter $a_{1n}(t)$. We then
find
\begin{align}
\Theta &\approx \theta \frac{ \int_0^\infty f_0(q)\di q q^3 \frac{q^2}{\epsilon^2} \para{\frac{q}{\epsilon}}^n}
	{\int_0^\infty f_0(q)\di q q^3 \para{\frac{q}{\epsilon}}^n }.
\end{align}
The guess $c_\text{vis}^2 = \frac{3}{4}w\para{1+w}$ can be seen to be
a special case of this approach having $n=-1$.  The problem is that
the value of $n$ best approximating the behavior of $\Psi_1$ and other
momenta is not the same in the relativistic and non-relativistic
limit.  In fact our testing shows that this guess sources $\sigma$ too
much during the relativistic to non-relativistic transition compared
to the exact solution. Instead we got much better results by using
$c_\text{vis}^2 = 3wc_g^2$, which avoids this excessive sourcing
during the transition, while still reducing to $1/3$ in the
relativistic limit.

For the ratio $\Sigma/\sigma$, the assumption of a $q$-independent 
$\Psi_2$ (i.e. $n=0$) yields $w_\sigma=
\mathfrak{p}/(3p)$, which provides satisfactory results and is adopted 
in the schemes described below. 

We speculate
that by pushing these kinds of considerations further, one could find
better approximations for $c_\text{eff}^2$, $c_\text{vis}^2$ and $w_\sigma$. It is also possible that another independent equation could be found, and that it would allow a better determination of $c_\text{eff}$.

\subsection{Implementation in CLASS}

For comparison, we have implemented 3 different Non-Cold Dark Matter
Fluid Approximations (NCDMFA) in \CLASS{} which differ only in their
respective equation for the shear. In correspondence with the
Ultra-relativistic Fluid Approximation discussed
in~\cite{class_approx}, we have named the approximations {\tt MB},
{\tt Hu} and {\tt CLASS}: in the relativistic limit,
they reduce to their relativistic counterpart in~\cite{class_approx}.
In all three
approximations we are using Eq.~\eqref{eq:continuity}
and~\eqref{eq:Euler} as the first two equations with
$c_\text{eff}=c_g$. The respective equations for the shear read
\begin{subequations}
\begin{align}
\dot{\sigma}_\text{\tt{MB}} &=-3\para{\frac{1}{\tau}+ \frac{\dot{a}}{a}\bpara{\frac{2}{3}-c_g^2-\frac{1}{3}\frac{\mathfrak{p}}{p}}}\sigma + \frac{4}{3} \frac{c_\text{vis}^2}{1+w}\bpara{2\theta + \dot{h} + 6\dot{\eta}}, &c_\text{vis}^2 = 3wc_g^2, \label{eq:shear_MB}\\
\dot{\sigma}_\text{\tt{Hu}} &=-3\frac{\dot{a}}{a} \frac{c_g^2}{w}\sigma + \frac{4}{3} \frac{c_\text{vis}^2}{1+w}\bpara{2\theta + \dot{h} + 6\dot{\eta}}, &c_\text{vis}^2 = w, \label{eq:shear_Hu2} \\
\dot{\sigma}_\text{\tt{CLASS}} &=-3\para{\frac{1}{\tau}+ \frac{\dot{a}}{a}\bpara{\frac{2}{3}-c_g^2-\frac{1}{3}\frac{\mathfrak{p}}{p}}}\sigma + \frac{4}{3} \frac{c_\text{vis}^2}{1+w}\bpara{2\theta + \dot{h} }, &c_\text{vis}^2 = 3wc_g^2. \label{eq:shear_CLASS}
\end{align}
\end{subequations}
The second shear equation, named {\tt Hu}, corresponds exactly to the prescription of
Ref.~\cite{Hu:1998kj} for approximating massive neutrinos.  The first
shear equation, {\tt MB}, comes directly from~\eqref{eq:shear_exact_syn} with the values
of $w_\sigma$ and $c_\text{vis}$ motivated in the previous subsection.
Finally, in~\cite{class_approx}, we found that removing the
$\dot{\eta}$ term leads to slightly better results for the matter
power spectrum, and can be justified using an analytic approximation
to the exact equations. By analogy, we also define in the massive
neutrino case a {\tt CLASS} approximation identical to the {\tt MB} one
except for the omission of this term.

\FIGURE{
%
\includegraphics[width=0.48\columnwidth]{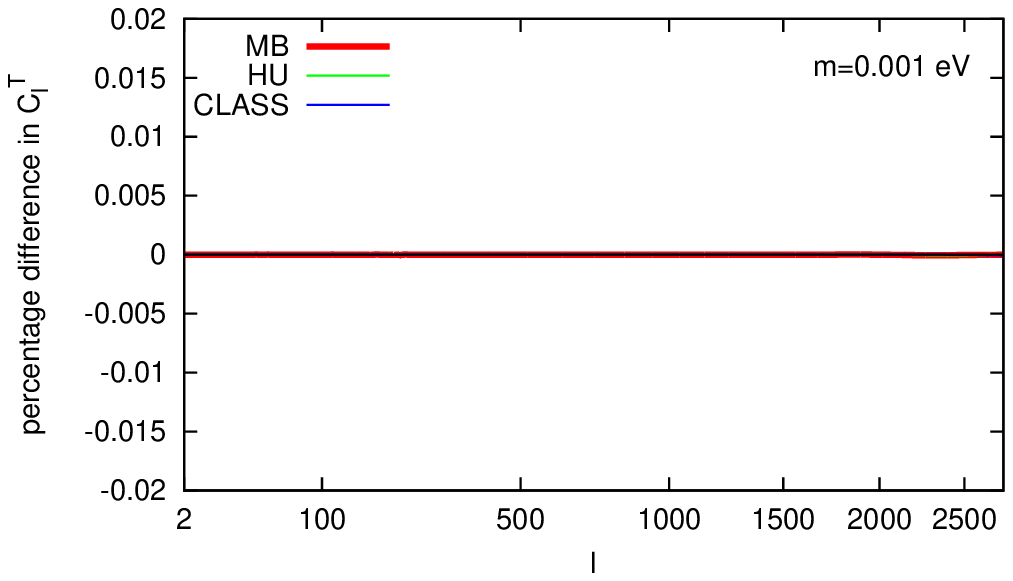}
\includegraphics[width=0.48\columnwidth]{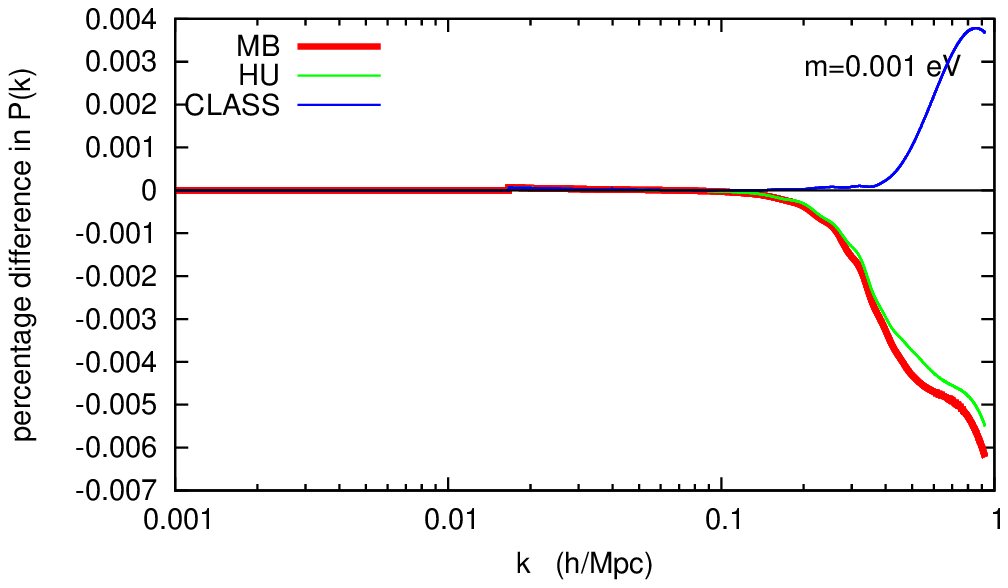} \\
\includegraphics[width=0.48\columnwidth]{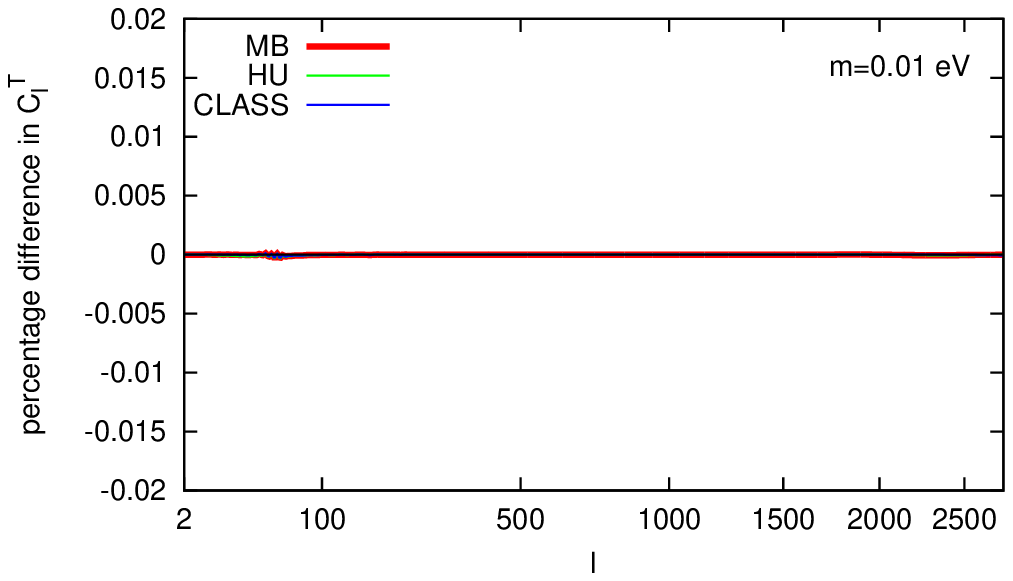}
\includegraphics[width=0.48\columnwidth]{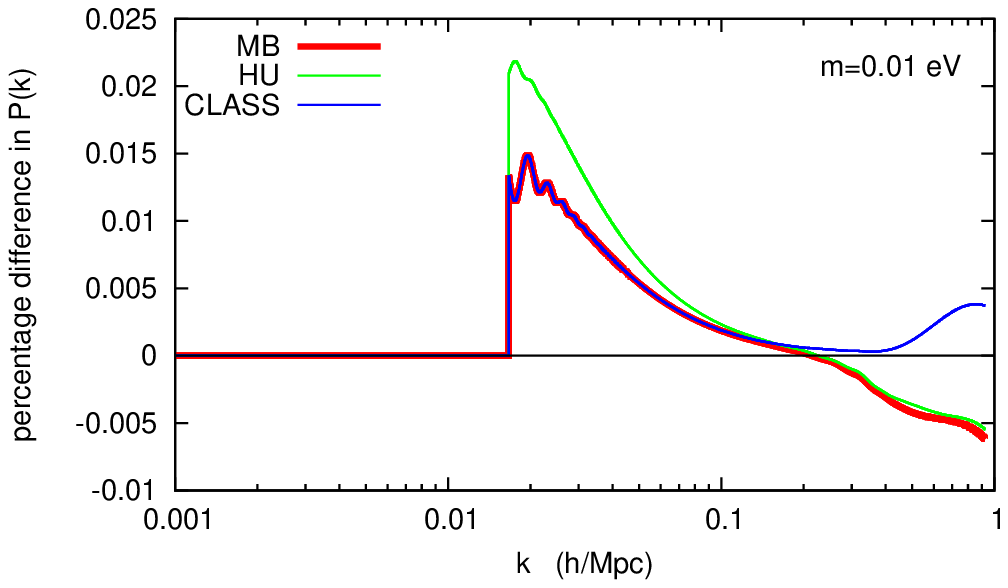} \\
\includegraphics[width=0.48\columnwidth]{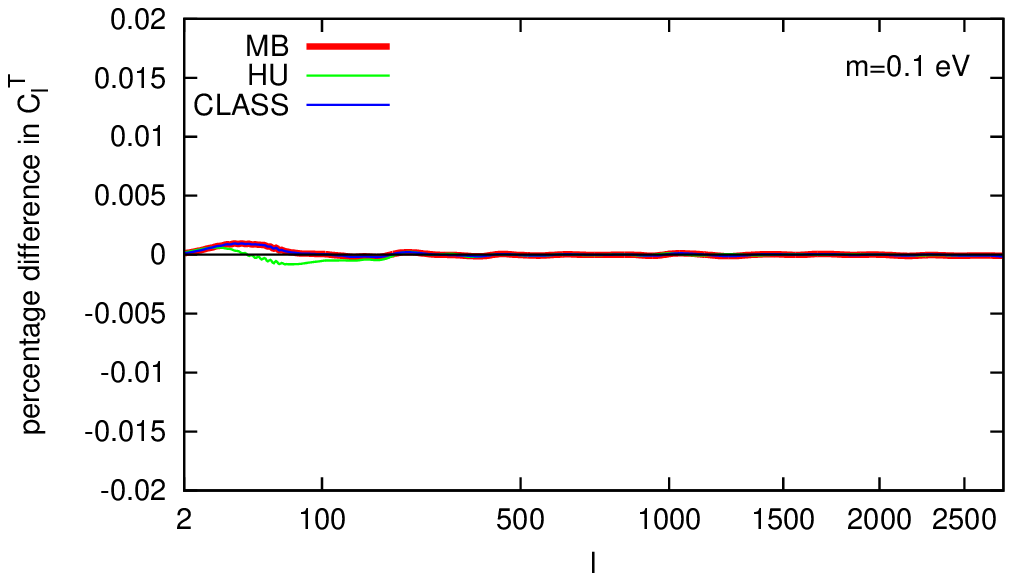}
\includegraphics[width=0.48\columnwidth]{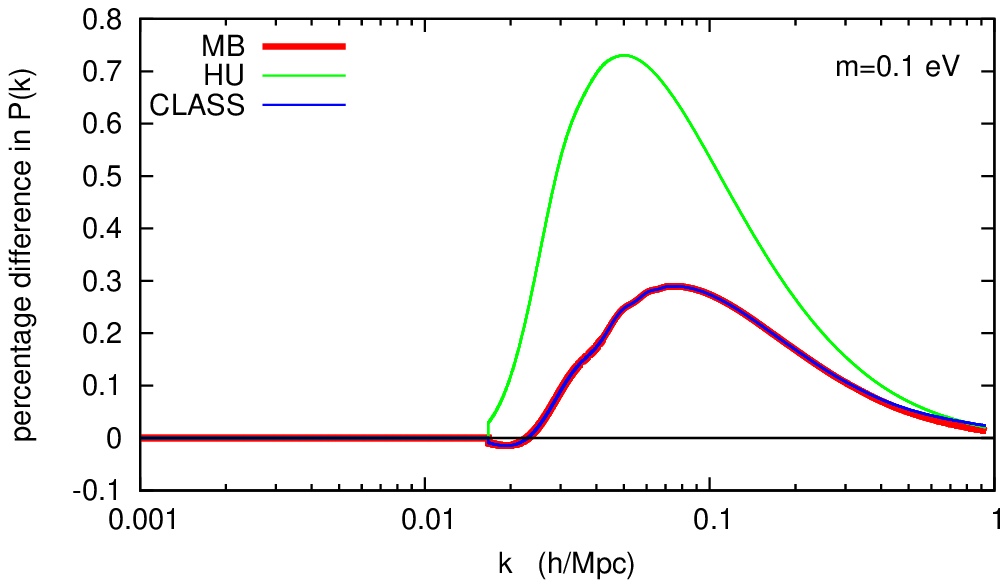} \\
\includegraphics[width=0.48\columnwidth]{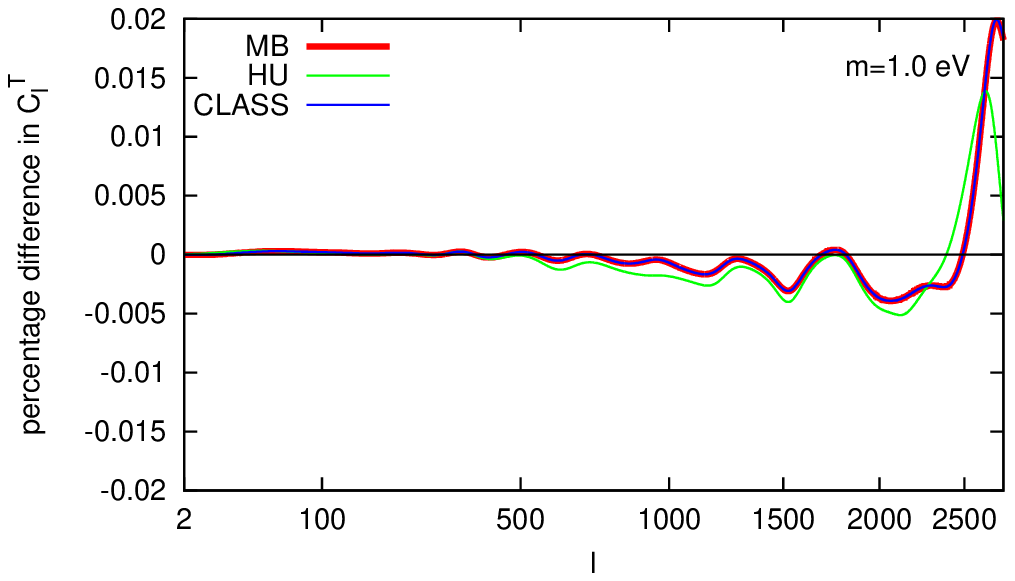}
\includegraphics[width=0.48\columnwidth]{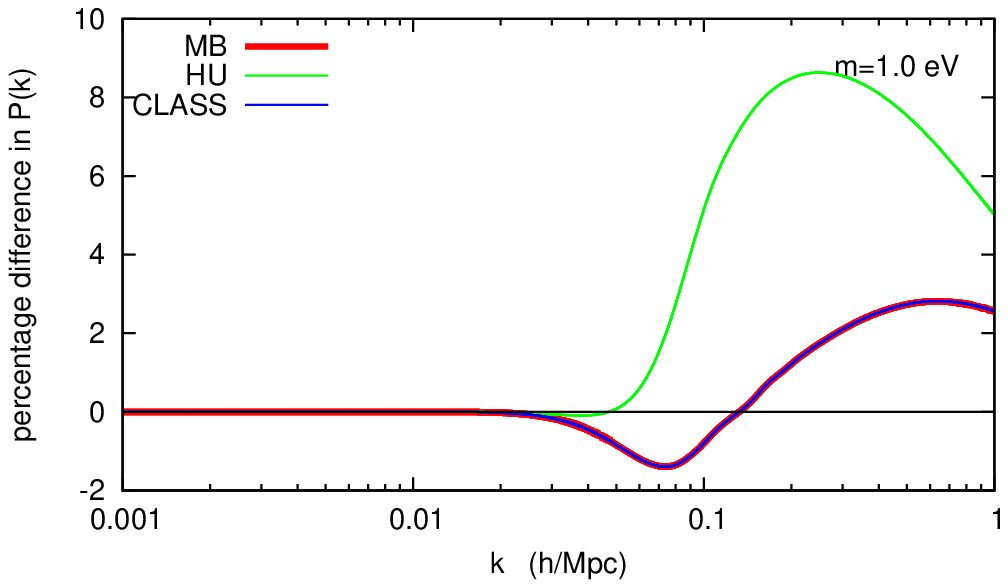} \\
\caption{\label{fig:fluid} On the left, we have shown the percentage
difference in the $C_l^T$ for three degenerate neutrino species with
mass $m=0.001\eV{}$, $m=0.01\eV{}$, $m=0.1\eV{}$ and $m=1\eV{}$
respectively, in runs with/without the fluid approximation. The fluid
approximation works very well as long as the neutrinos are
relativistic, so this is what we expect. On the right we have shown
the matter power spectrum for the same masses. Here the agreement is
not so good as the mass becomes higher.}
}
In Fig.~\ref{fig:fluid} we have tested
these three fluid approximations in a model with no massless neutrinos
and 3 degenerate massive neutrinos. The three approximations work very well as long as the neutrinos are
light and become non-relativistic after photon decoupling. Like in the
massless case, the \CLASS{} approximation is slightly better for
predicting the matter power spectrum on small scales, and we set it to
be the default method in the code. When the mass increases, the fluid
approximation alters the CMB spectra on small angular scales ($l \geq
2500$), but the error remains tiny (only 0.02\% for $l=2750$ for three
species with $m=1\eV{}$). The effect on the matter power spectrum is
stronger: with three 1 \eV{} neutrinos, the $P(k)$ is wrong by 1 to 3\%
for $k \in [0.05; 1]h\text{Mpc}^{-1}$. Hence, we recommend to use the
fluid approximation for any value of the mass when computing CMB
anisotropies, and only below a total mass of one or two \eV{}'s when
computing the matter power spectrum. However, cosmological bounds on
neutrino masses strongly disfavour larger values of the total mass. This
means that in most projects, \CLASS{} users can safely use the fluid
approximation for fitting both CMB and large scale structure data.

\section{Standard massive neutrinos \label{neutrinos}}
We first illustrate our approach with the simple case of standard
massive neutrinos with a Fermi-Dirac distribution. In this case, for
each neutrino, the user should provide two numbers in the input file:
the mass $m$, and the relative temperature {\tt T\_ncdm}~$\equiv
T_\nu/T_\gamma$ (the ratio of neutrino to the photon temperature).
The \CLASS{} input file {\tt explanatory.ini} recommends to use the value
{\tt T\_ncdm}=0.71599, which is ``fudged'' in order to provide a
mass-to-density ratio $m/\omega_\nu = 93.14$~\eV{} in the
non-relativistic limit. This number gives a very good approximation to
the actual relic density of active neutrinos, resulting from an
accurate study of neutrino decoupling~\cite{Mangano:2005cc}.  However,
when comparing the \CLASS{} results with those from \CAMB{}, we take {\tt
T\_ncdm}=0.7133 in order to recover the mass-to-density ratio assumed
in that code. Finally, if no temperature is entered, the code will default to
the instantaneous decoupling value of $\para{4/11}^{1/3}$.
\subsection{Agreement with CAMB}
\FIGURE{
%
\includegraphics[width=0.48\columnwidth]{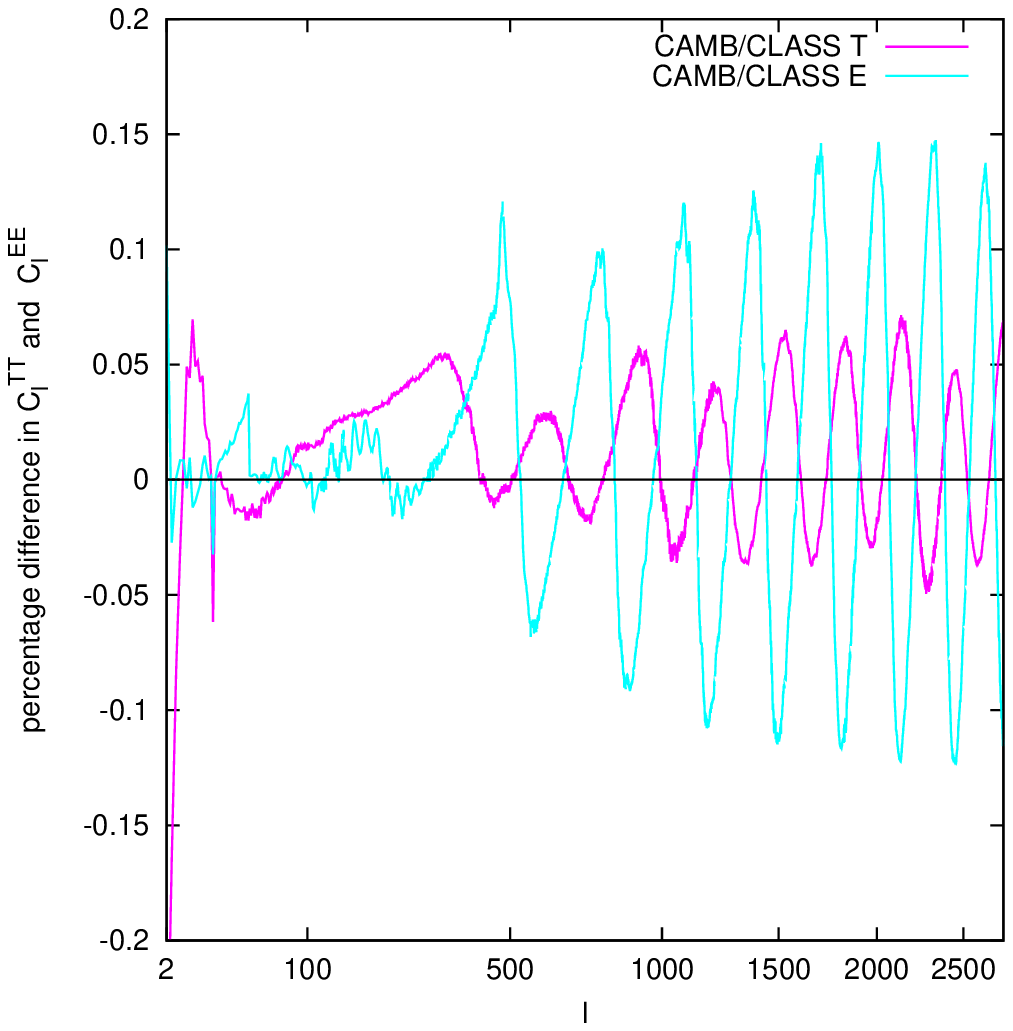}
\includegraphics[width=0.48\columnwidth]{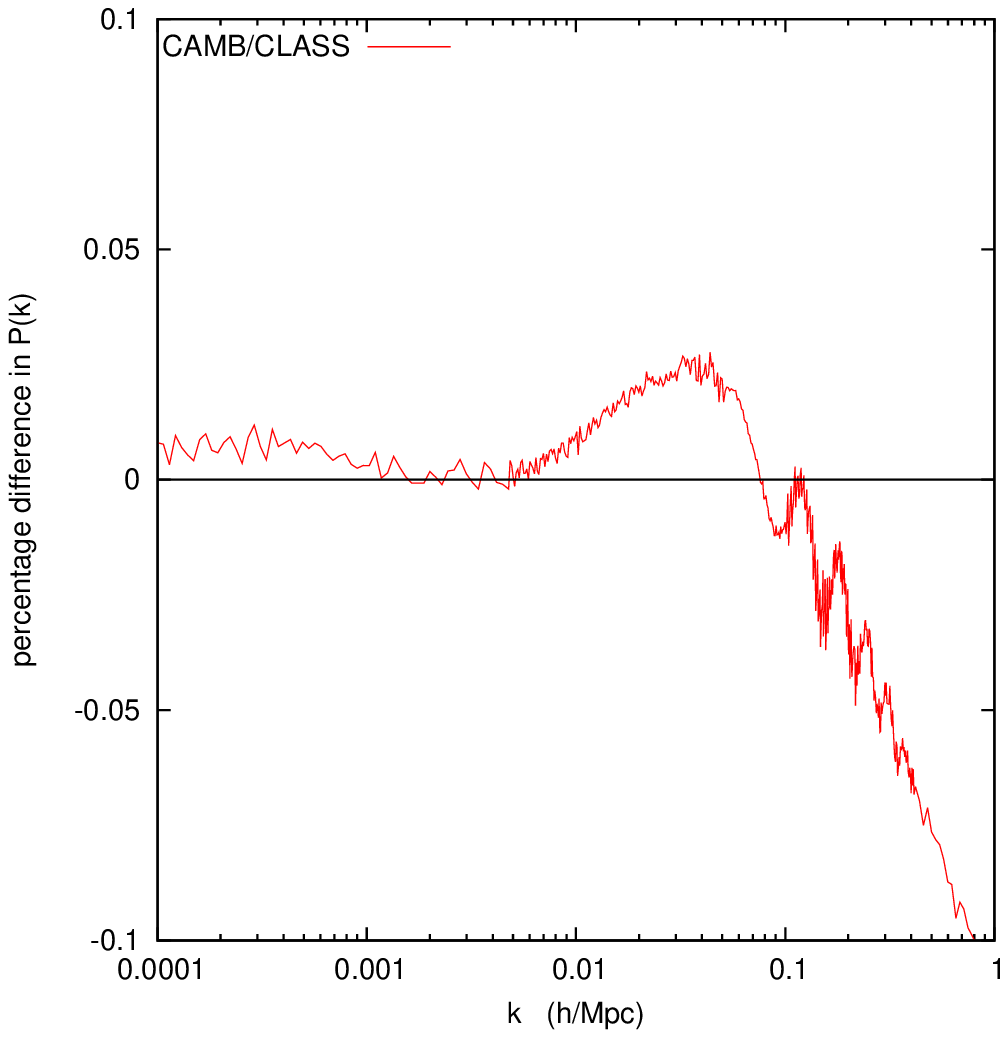} \\
\caption{\label{fig:compare_camb} Relative difference between \CAMB{} and \CLASS{} spectra in a model with $\Omega_\nu = 0.02$, two massless neutrinos, and reference accuracy settings. The two codes agree rather well.}
}
\FIGURE{
%
\includegraphics[width=0.48\columnwidth]{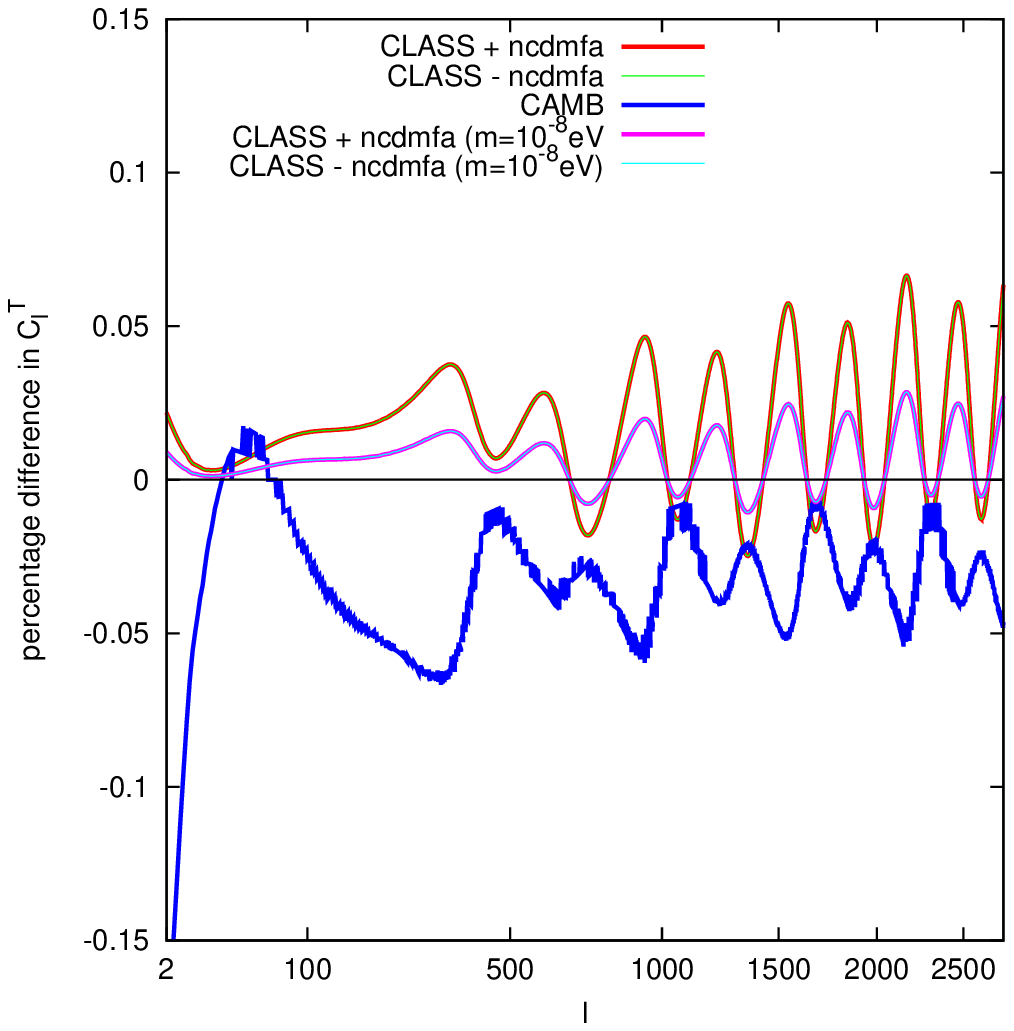}
\includegraphics[width=0.48\columnwidth]{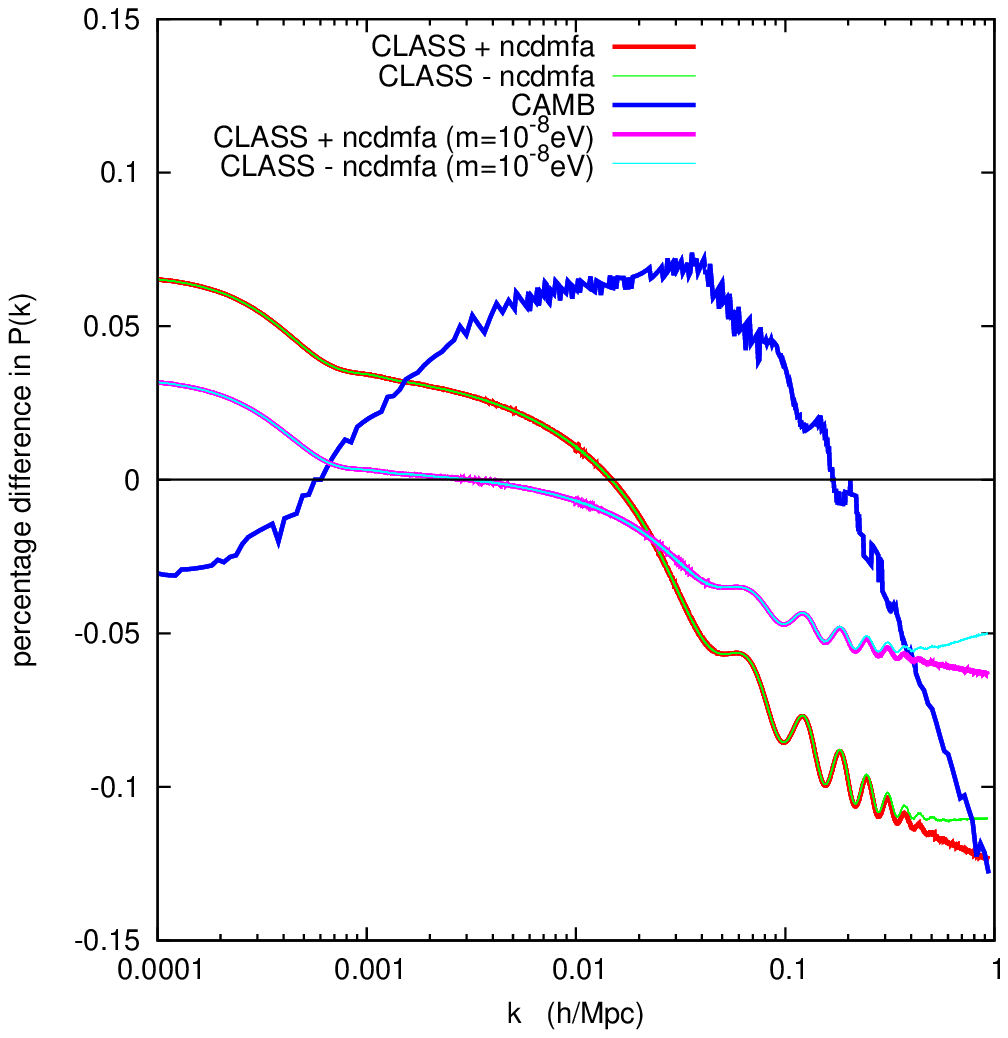} \\
\caption{\label{fig:massless_limit} This is a test of how well \CAMB{} and \CLASS{} recovers the massless limit. We compute a model with $\Omega_\nu = 1.2\cdot 10^{-4}$ and 3 massive neutrinos with degenerate mass. This setting corresponds to a neutrino mass of $m_i=2.8\cdot 10^{-4}\eV{}$, which is not exactly massless, but it is the best we can do since the mass parameter can not be set directly in \CAMB{}. Setting the mass parameter in \CLASS{} to $m_i=10^{-8}\eV{}$ reveals that we are in part seeing the effect of the neutrino going slightly non-relativistic at late times.}
}
In Fig.~\ref{fig:compare_camb}, we compare the CMB and matter power
spectrum from \CAMB{} and \CLASS{} (without the NCDM fluid approximation)
for two massless and one massive neutrino with $\Omega_\nu=0.02$
(corresponding to a mass $m \simeq 0.923 \eV{}$). We used high accuracy
settings for \CAMB{}, described in \cite{class_comp} under the name {\it
[CAMB:07]}.  For \CLASS{}, we used the input file {\tt cl\_ref.pre},
which corresponds to the setting {\it [CLASS:01]} in \cite{class_comp}
for parameters not related to NCDM; for the latter, {\tt
cl\_ref.pre} contains the settings described in the first column of
Table~\ref{settings}.  For such settings and in absence of massive
neutrinos, the two temperature spectra would agree at the 0.01\% level
in the range $l \in [20;3000]$; at the 0.02\% level for polarization in the
same range; and at the 0.01\% level for the matter power spectrum for
$k<1h\text{Mpc}^{-1}$.  With a neutrino mass close to 1~\eV{}, we see in
Fig.~\ref{fig:compare_camb} that the discrepancy is approximately
six times larger than in the massless case. However, it remains very
small: even with massive neutrinos the two codes agree to better than
0.1\% for the CMB and matter power spectra. This is by far sufficient
for practical applications.

In a perfect implementation of massless and massive neutrinos in
Boltzmann codes, we expect that in the relativistic limit $m \ll
T_\nu^0$ (where $T_\nu^0$ is the neutrino temperature today) the
spectra would tend towards those obtained with three massless species
(provided that we are careful enough to keep the same number of
relativistic degrees of freedom $N_{\rm eff}$). We performed this
exercise for both codes, and the results are presented in
Fig.~\ref{fig:massless_limit}. It appears that with a small enough
mass, \CLASS{} can get arbitrarily close to the fully relativistic case:
with a mass of $10^{-8} \eV{}$, the difference is at most of 0.03\% in
the $C_l$'s and 0.05\% in the $P(k)$. This test is another way to validate
the accuracy of our implementation.

\subsection{Precision files}

We now come to the question of defining degraded accuracy settings
for computing the spectra in a fast way, while keeping the
accuracy of the results under control. For such an exercise, we need
to define a measure a precision. Like in \cite{class_comp}, we will
use an effective $\chi^2$ which mimics the sensitivity of a CMB
experiment like Planck to temperature and E-polarisation anisotropies.

This is a time consuming exercise, which we have described in details in Sec.~\ref{accuracy} of the appendix. The end result is a set of precision-files which guarantees a certain precision when used, and these files are of course available for download on the \CLASS{} web page\footnote{{\tt http://class-code.net}}.

\subsection{Performance}
The quadrature method reveals to be extremely useful since even with
five values of the momenta, we get accurate results leading to
0.2\%-0.3\% accuracy on the $C_l's$, 0.1\% accuracy on the $P(k)$ and
$\Delta \chi^2 \sim 1$.  Traditional Boltzmann codes employ 14 momenta
in order to achieve a comparable precision. In the presence of massive
neutrinos, the total execution time of a Boltzmann code is dominated
by the integration of the perturbation equations, which depends on
the total number of perturbed variables, itself dominated by the
number of massive neutrino equations. By reducing the number of
momenta from 14 to 5, the quadrature method speeds up the code by more
than a factor two. We find that the use of the fluid approximation leads to an
additional 25\% speed up for standard accuracy settings (like those in
the file {\tt chi2pl1.pre}). In total, for a single massive neutrino,
our method speeds up the code by a factor 3. This means that instead
of being 4.5 times slower in presence of one massive neutrino, \CLASS{}
only becomes 1.5 times slower. We checked these numbers with various masses
and accuracy settings.
\subsection{Realistic mass schemes}

\FIGURE{
%
\includegraphics[width=0.96\columnwidth]{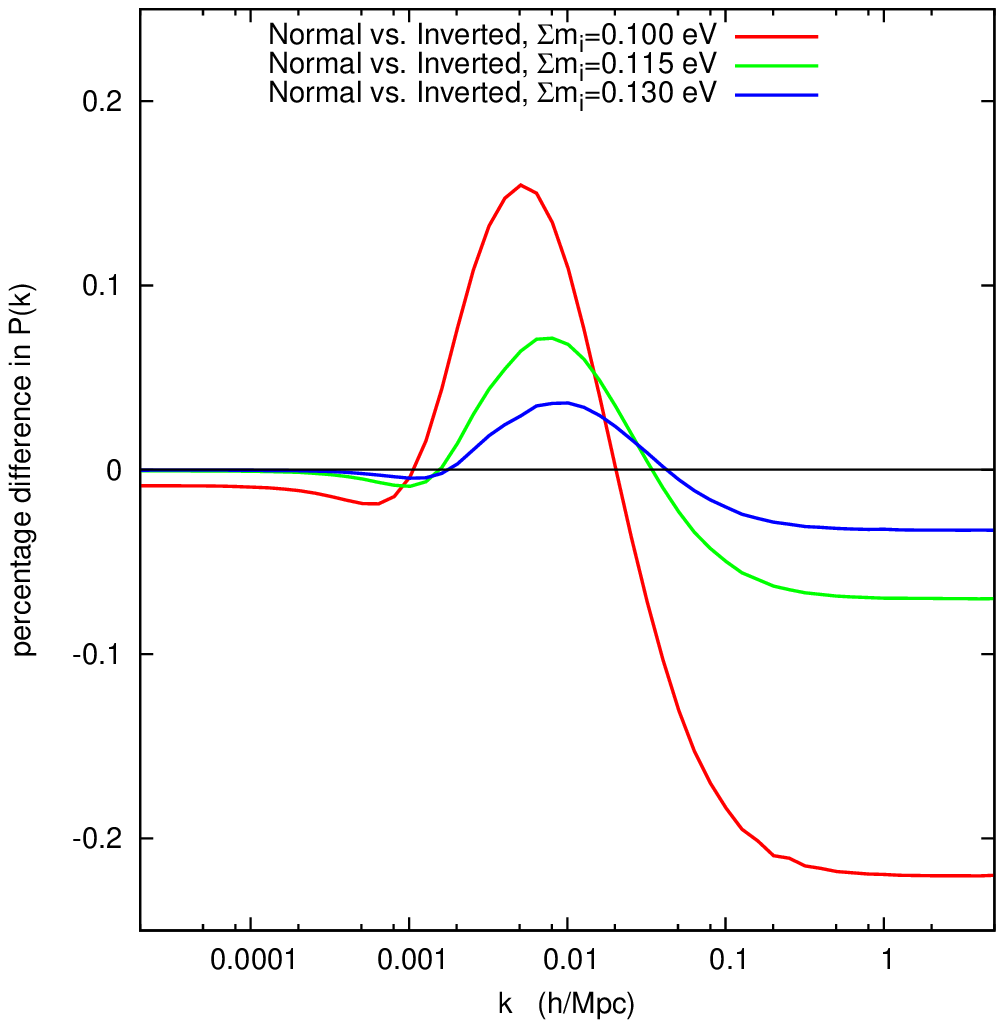}
\caption{\label{fig:normal_vs_interted} Ratio of matter power spectra
for pairs of models with three massive neutrinos, obeying either to
the normal or inverted hierarchy scenario, but with a common total
mass for each pair: $M_\nu=0.100$~eV, $0.115$~eV or $0.130$~eV. The various effects
observed here are discussed in the text. }
}

We have proved in this section that \CLASS{} can be employed in any
project requiring high-precision computations of cosmological
observables in presence of massive neutrinos. It is of course
perfectly suited for realistic situations with different neutrino
species and masses. 

To illustrate this, we display in Figure~\ref{fig:normal_vs_interted}
the ratio of pairs of matter power spectra for models with three
massive neutrinos satisfying constraints from atmospheric/solar
oscillation experiments~\cite{Nakamura:2010zzi} ($\Delta m^2_{21}=7.6\times10^{-5}{\rm eV}^2$,
$\Delta m^2_{32}=\pm 2.4\times10^{-3}{\rm eV}^2$).  Each pair of models
corresponds to one normal hierarchy and one inverted hierarchy scenario,
with the same total mass $M_\nu$, equal to 
$0.100$~eV, $0.115$~eV or $0.130$~eV. The first total mass is very close to the minimum allowed value for the inverted hierarchy, $M_\nu\simeq 0.0994$~eV.
For each pair or models with a given $M_\nu$:
\begin{itemize} 
\item on intermediate scales, the bump reflects the
difference in the three free-streaming scales involved in the two
models. 
\item 
in the large $k$ limit, the two spectra are offset by 0.03\% to
0.22\%: it is known that in this limit, the suppression in the power
spectrum induced by neutrino free-streaming depends mainly on the
total mass (through the famous $-8f_{\nu}$ approximate formula), but
also slightly on the mass splitting (in \cite{Lesgourgues:2006nd}, a
more accurate formula gives the suppression as a function of both the
total mass and number of degenerate massive neutrinos). When $M_\nu$
increases, the two models are less different from each other (they go
towards a common limit, namely the degenerate mass scenario), and the
discrepancy is less pronounced.
\item 
in the small $k$ limit, the two spectra are nearly identical. The tiny
difference, which increases when $M_\nu$ decreases, is due to the fact
that in the inverted hierarchy model, there is a very light neutrino
just finishing to complete its non-relativistic transition today. It therefore has a non-negligible pressure, which slightly affects metric
perturbations on large wavelengths.
\end{itemize}
Observing the difference between these two models would be extremely
challenging, although 21 cm surveys could reach enough sensitivity
\cite{Pritchard:2009zz}.

\section{Beyond standard massive neutrinos\label{exemples}}
In this section we will illustrate the power and flexibility of the non-cold Dark Matter implementation in \CLASS{}, by implementing different models which have already been studied elsewhere in the literature.
\subsection{Massive neutrinos with large non-thermal corrections}
It is plausible that some new physics can introduce non-thermal
corrections to an otherwise thermal Fermi-Dirac distribution function. One
might think of using CMB and large scale structure data to put bounds on
such non-thermal corrections, as was described
e.g. in~\cite{Cuoco:2005qr}.  \CLASS{} is ideally suited for playing with
such models. As a test case, we take the following distribution
from~\cite{Cuoco:2005qr}:
\begin{align}
f(q) &= \frac{2}{\para{2\pi}^3}\bpara{\frac{1}{e^q+1} + \frac{A\pi^2}{q^2\sqrt{2\pi}\sigma} \exp\para{-\frac{\para{q-q_c}^2}{2\sigma^2}} },
\label{eq:peak}
\end{align}
which is the Fermi-Dirac distribution with an added Gaussian peak in
the number density. This distribution could presumably be the result
of some particle suddenly decaying into neutrinos at a late time.

In practise, we only need to change the expression for $f(q)$ in
\CLASS{}, which appears in a unique line (in the function {\tt
background\_ncdm\_distribution()}). All the rest, like density-to-mass relation and computation of the logarithmic derivative, is done automatically
by the code. In particular, we do not need to change the accuracy
parameters {\tt tol\_ncdm} and {\tt tol\_ncdm\_bg}: the momentum
sampling algorithm automatically increases the number of momenta by a
significant amount, in order to keep the same precision. If this was not the case, the effect of the peak would be underestimated because of under sampling, and the parameter extraction would then likely be biased.
\FIGURE{
%
\includegraphics[width=0.48\columnwidth]{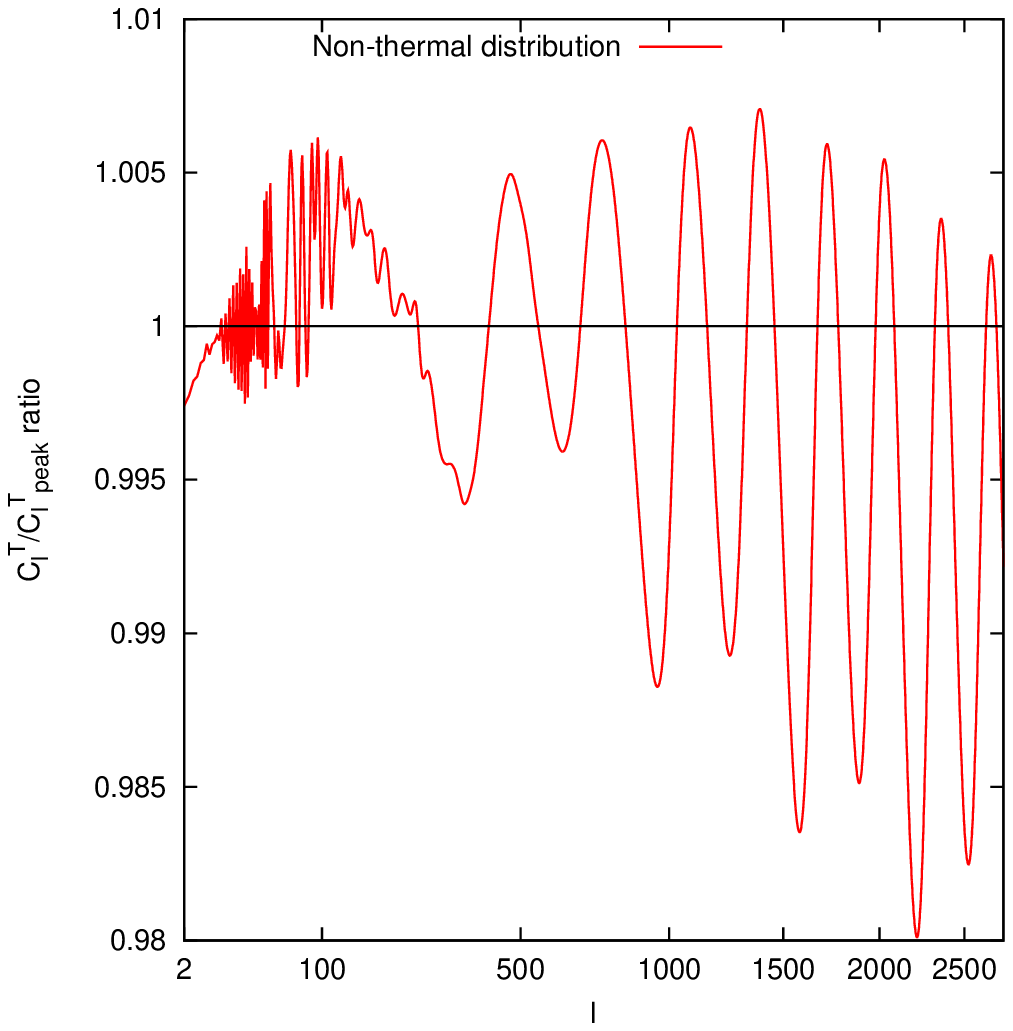}
\includegraphics[width=0.48\columnwidth]{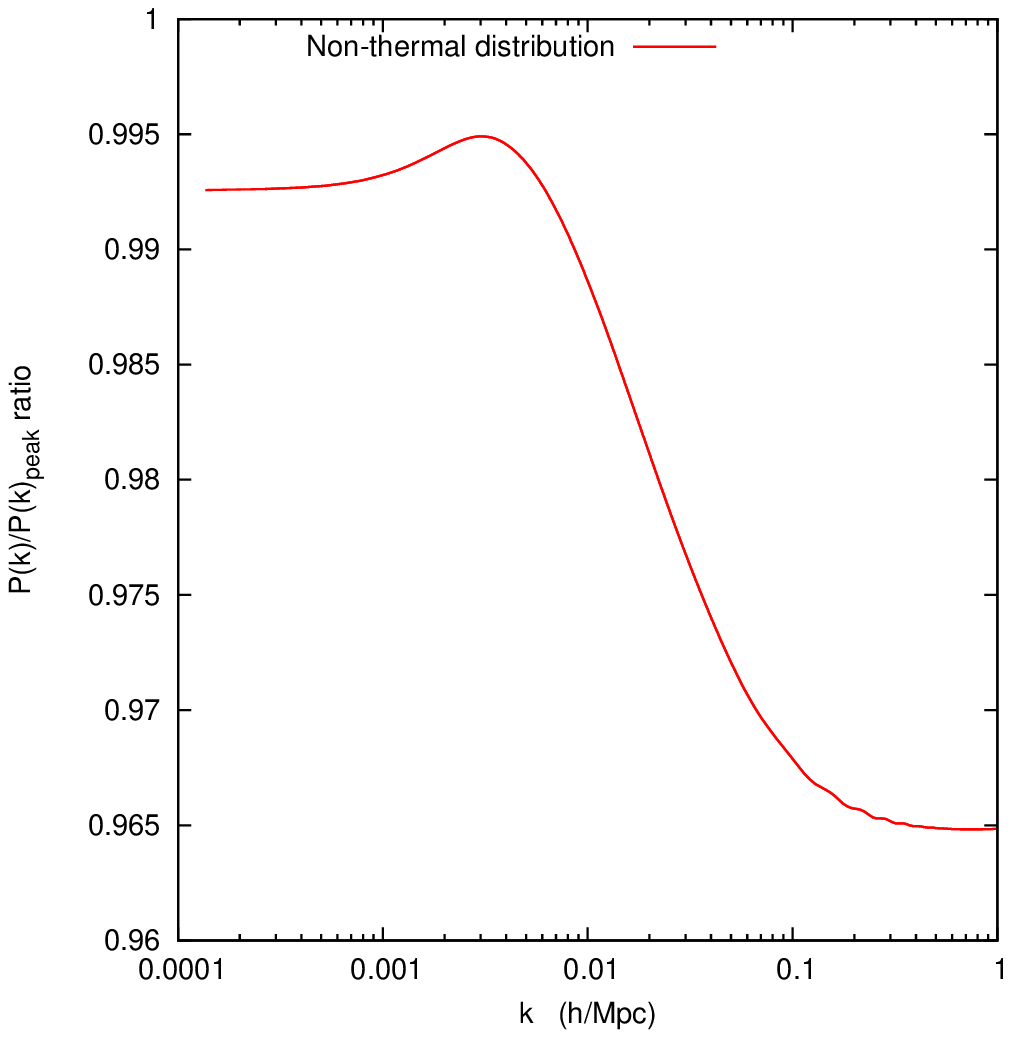}
\caption{\label{fig:peak} $C_l$'s and $P(k)$'s for a model of 3
degenerate neutrinos with the non-thermal distribution~\eqref{eq:peak}
using parameters $m=1.0$~eV, $A=0.018$, $\sigma=1.0$ and $q_c =
10.5$. This corresponds to $N_\text{eff} = 3.98486$. We have compared
this model to a model with degenerate thermal neutrinos with the same
mass and $N_\text{eff}$.
The signal is due to a combination of
background and perturbation effects: although the mass and the relativistic density are
the same, the non-relativistic density and the average momentum differ
significantly in the two models.
}}

In Fig.~\ref{fig:peak}, we show the CMB and matter power spectra for
this model, relative to a standard model with three thermally
distributed neutrinos. The two models are chosen to share exactly the
same masses and the same initial number of relativistic degrees of
freedom $N_\text{eff}$. Nevertheless, they do not have the
same non-relativistic neutrino density and average neutrino momentum;
in particular, non-thermal neutrinos in the decay peak become
non-relativistic slightly later. This induces a combination of
background and perturbation effects affecting CMB and matter power
spectra in a significant way.

\subsection{Warm dark matter with thermal-like distribution}

There is an infinity of possible warm dark matter models, since the
phase-space distribution of warm dark matter depend on the details of
its production mechanism. The most widely studied model is that of
non-resonantly produced warm dark matter with a rescaled Fermi-Dirac
distribution, having the same temperature as that of active neutrinos.
This model is implemented in the default \CLASS{} version: when the user
enters a temperature, a mass and a density $\Omega_{\rm ncdm}$ (or $\omega_{\rm ncdm}$) 
for the same species, the code knows that the
degeneracy parameter in front of the Fermi-Dirac distribution must be
rescaled in order to match these three constraints simultaneously.
The code will also ensure that the perturbations begin to be
integrated when the non-cold species is still relativistic, in order
to properly follow the transition to the non-relativistic regime.
\FIGURE{
%
\includegraphics[width=0.48\columnwidth]{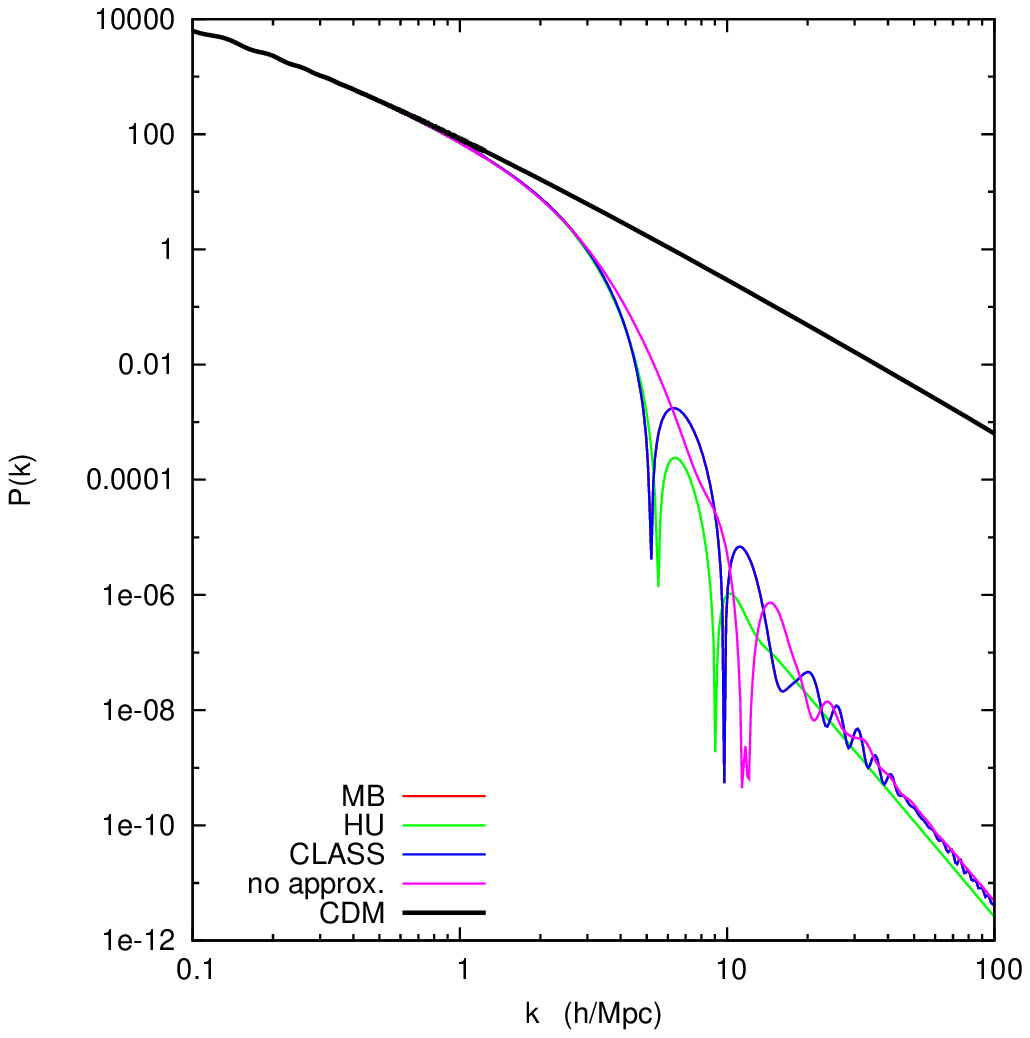}
\includegraphics[width=0.48\columnwidth]{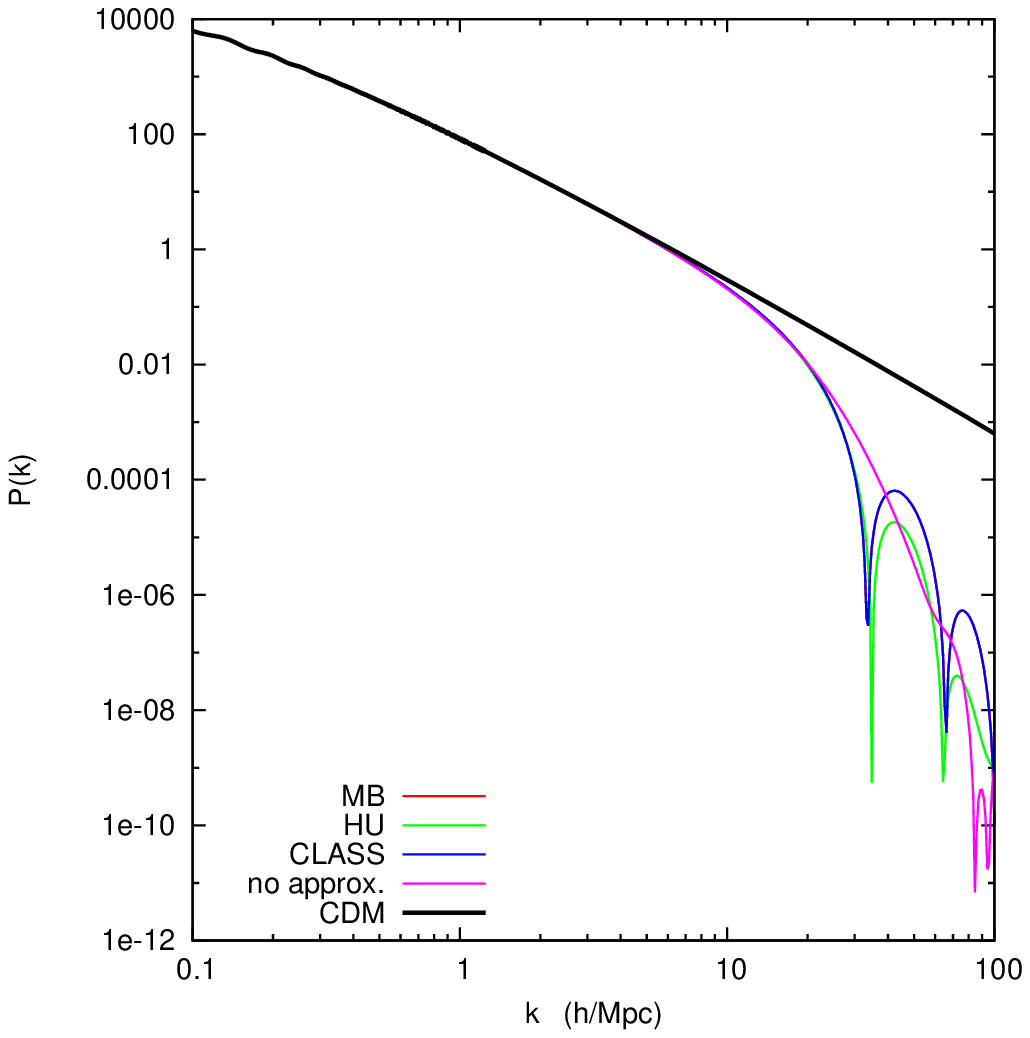}
\caption{\label{fig:wdm} $P(k)$'s for a Warm Dark Matter model
with $m=1$~keV (left) and $m=10$~keV (right). The fluid
approximation can be seen to be a very good approximation in this
case, though it does not catch the acoustic oscillations precisely. }
}

We illustrate this by running a $\Lambda$WDM model with a mass of
$m=1\text{keV}$ or
$m=10\text{keV}$ and a density $\Omega_{\rm ncdm}=0.25$, with or
without the fluid approximation. We compare the results with those of
$\Lambda$CDM with $\Omega_{\rm cdm}=0.25$, in order to show the
well-known suppression effect of WDM in the small-scale limit of the
matter power spectrum.
It appears that the fluid approximation works very well in those cases,
unless one wants to resolve the details of the WDM acoustic oscillations
on very small scales, first predicted in~\cite{Boyanovsky:2010pw}.
\subsection{Warm dark matter with non-trivial production mechanism}
Non-resonantly produced warm dark matter candidates are severely
constrained by Lyman-$\alpha$ bounds, but such bounds do not apply to
other warm particles which could have been produced through more
complicated mechanisms (e.g. resonant production), leading to a
non-trivial, model-dependent phase-space distribution
function~\cite{Boyarsky:2008mt}. It is not always easy to find a good
analytic approximation for such a distribution; this is anyway not an
issue for \CLASS{}, since the code can read tabulated values of
$f(p)$ from an input file.
\FIGURE{
%
\includegraphics[width=0.48\columnwidth]{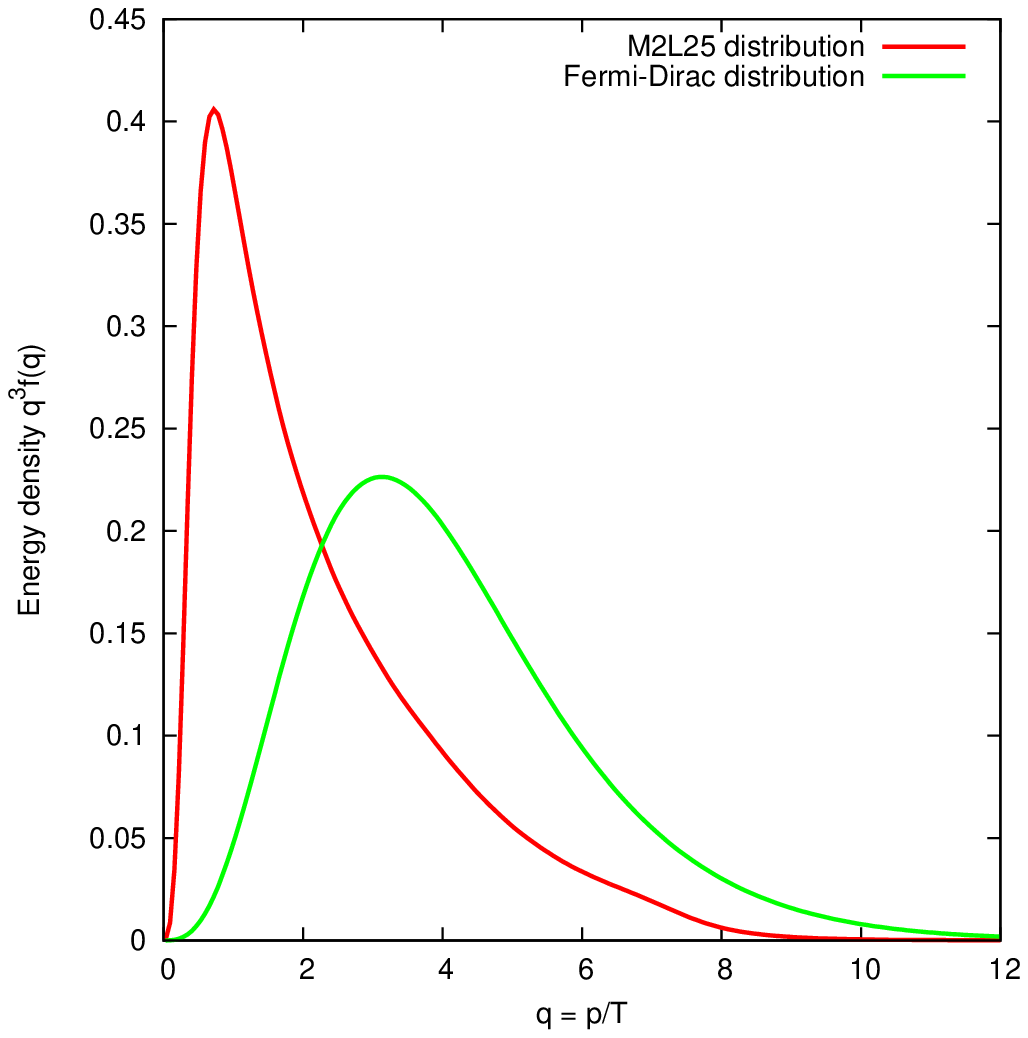}
\includegraphics[width=0.48\columnwidth]{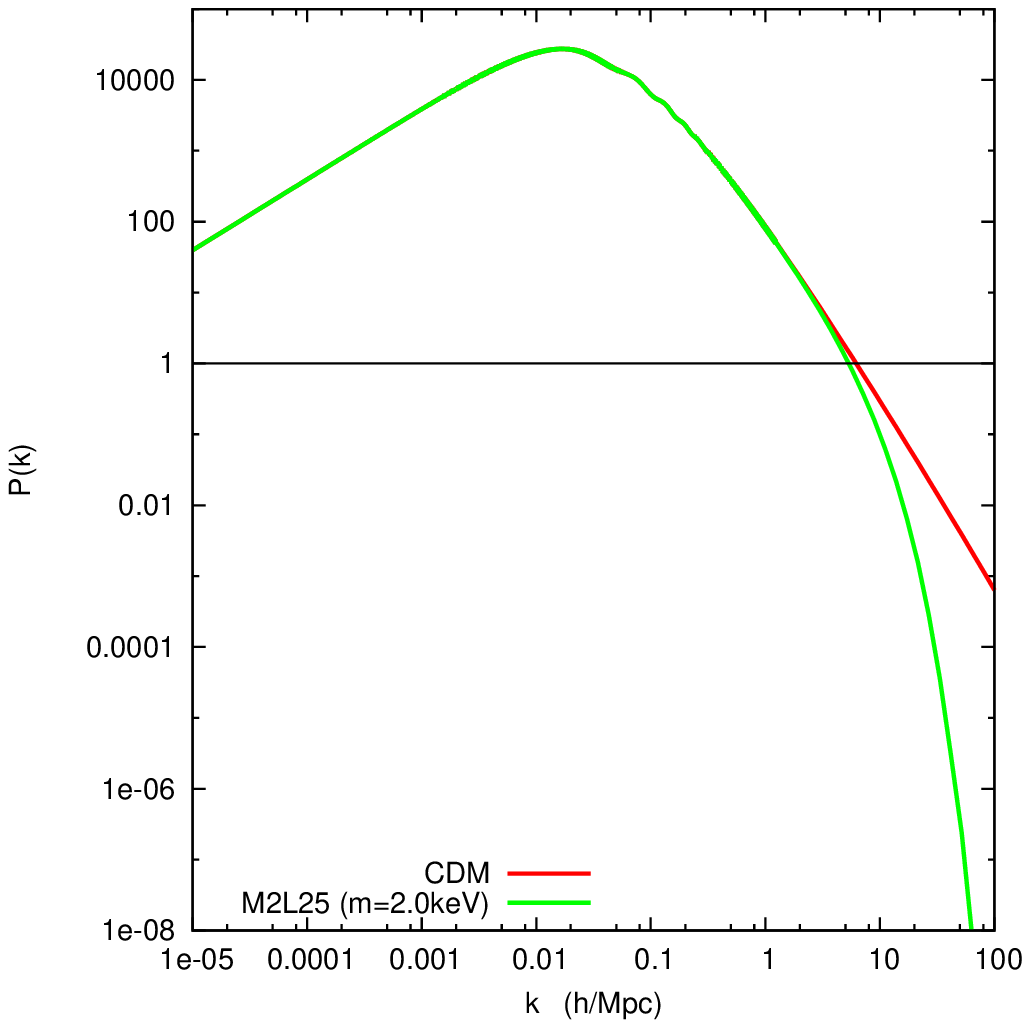}
\caption{\label{fig:M2L25} $P(k)$'s for a Warm Dark Matter model with a non-trivial production mechanism for a mass of $m=2$~keV compared to the same model with Cold Dark Matter. Note that the normalisation of the distribution function is arbitrary; when both {\tt m\_ncdm} and one of \{{\tt Omega\_ncdm,omega\_ncdm}\} is present for some species, \CLASS{} will normalise the distribution consistently.} 
}

We illustrate this case by taking a particular model for resonantly
produced sterile neutrinos, which distribution was computed
numerically by \cite{Laine:2008pg} (simulating the details of sterile
neutrino production and freeze-out), and stored in a file with
discrete $q_i$, $f_i$ values.  Again, we only need to specify the name
of this file in the \CLASS{} input file, to enter a value for the mass
and for the density $\Omega_{\rm ncdm}$, and the rest is done
automatically by the code (finding the mass-density relation and the
correct normalization factor for $f(q)$, defining the new momentum
steps, deriving $[d \ln f]/[d \ln q]$ with a good enough accuracy). The
assumed $f(q)$ and the resulting matter power spectrum when the mass
is set to $m=2\text{keV}$ is shown in Fig.~\ref{fig:M2L25}. By eye,
this spectrum seems identical to a thermal-like WDM one, but the
cut-off is in fact much smoother due to an excess of low-momentum
particles in this model (which behave like a small cold dark matter fraction).

\section{Conclusions}

A large fraction of the activity in cosmology consists in deriving
bounds on particle physics in general, and on the neutrino and dark
matter sector in particular. Fitting cosmological data with non-standard neutrinos
or other non-cold relics require non-trivial changes in existing
public Boltzmann codes. Moreover, running parameter extraction codes
including massive neutrinos or more exotic non-cold relics is
computationally expensive due to a significant increase in the number
of differential equations to be solved numerically for each set of cosmological
parameters.

The newly released Cosmic Linear Anisotropy Solving System aims at
rendering this task easy and fast. The code provides a very
friendly and flexible input file in which users can specify a lot of
non-standard properties for the NCDM sector: masses, temperatures,
chemical potentials, degeneracy parameters, etc. Moreover, the
Fermi-Dirac distribution function is not hard-coded in \CLASS{}; it is
just a default choice appearing in one line of the code, which can be
very easily modified. Even when a non-thermal distribution $f(q)$ does
not have a simple analytic expression, the code can be told to read it
directly from a file. After reading this function, \CLASS{} performs a
series of steps in a fully automatic way: finding the mass-density
relation, defining an optimal sampling in momentum space with a
sophisticated but fast algorithm, and accurately computing the
derivative of $f(q)$, needed in the perturbation equations.

In this paper, we presented the main two improvements related to the
NCDM sector in \CLASS{}: an adaptive quadrature sampling algorithm,
which is useful both for the purpose of flexibility (the sampling is
always adapted to any new distribution function) and speed (the code
sticks to a minimum number of momenta, and hence, of perturbation
equations); and a fluid approximation switched on inside the Hubble
radius. We showed that the latter approximation works very well for
realistic active neutrinos (with a total mass smaller than $1-2\eV{}$), 
and for warm dark matter candidates becoming non-relativistic
during radiation domination. In between these two limits, there is a
range in which the accuracy of the fluid approximation is not well
established, and in which the user may need to keep the approximation off,
at the expense of increasing the execution time. However, the range
between a few eV and few keV is usually not relevant in most realistic
scenarios.

The adaptive quadrature sampling algorithm and the fluid
approximation both contribute to a reduction in the total execution
time of the code by a factor of three for ordinary neutrinos. This means
that when one massive neutrino species is added to the $\Lambda$CDM
model, \CLASS{} becomes 1.5 times slower instead of 4.5 times slower like other codes. Since 
the code is already quite fast in the the massless case, we conclude that the
global speed up is significant and appreciable when fitting
cosmological data.

\section*{Acknowledgments}
We wish to thank Oleg Ruchayskiy for providing the phase-space distribution of the Warm Dark Matter model, Fig.~\ref{fig:M2L25}, which gave us an opportunity to test the code in a non-standard scenario. We also wish to thank Emanuele Castorina for many useful discussions on the fluid approximation for neutrinos during his stay at CERN.

\appendix
\section{Optimal momentum sampling \label{sampling}}

We define a quadrature rule on $\mathcal{I}$ to be a set of weights $W_i$ and a set of nodes $q_i$, such that
\begin{align}
\mathcal{I} &\simeq \sum_{i=1}^{n} W_i g\para{q_i}.
\end{align}
Note that the distribution function itself has been absorbed into the
weights. The optimal quadrature rule will depend on both the
distribution $f_0(q)$ and the accuracy requirement {\tt tol\_ncdm},
but the specific method used for obtaining the rule is decoupled from
the rest of the code; the output is just two lists of $n$ points,
$\{q_i\}$ and $\{W_i\}$. \CLASS{} tries up to three different methods
for obtaining the most optimal quadrature rule, each with its own
strength and weaknesses. These are Gauss-Laguerre quadrature, adaptive
Gauss-Kronrod quadrature and a combined scheme. We will now
discuss each of them.

\subsection{Gauss-Laguerre quadrature}
Most of the time, the distribution function will be close to a Fermi-Dirac distribution, and the integrand is exponentially decaying with $q$. The Gauss-Laguerre quadrature formula is well suited for exponentially decaying integrands on the interval $(0;\infty)$, so this is an obvious choice. The rule is~\cite{AbramowitzStegun64}
\begin{align}
\int_0^\infty \di q e^{-q}h\para{q} &\simeq \sum_{i=1}^n w_i h\para{q_i},
\end{align}
where the nodes $q_i$ are the roots of $L_n$, the Laguerre polynomial of degree $n$ and the weights can be calculated from the formula
\begin{align}
w_i = \frac{q_i}{\para{n+1}^2\bpara{L_{n+1}\para{q_i}}^2}.
\end{align}
If we put $h(q) = e^q f_0(q) g(q)$ we obtain the rule
\begin{align}
W_i = w_i e^{q_i} f_0\para{q_i}.
\end{align}
This rule will be very effective when the ratio $f_0/e^{-q}$ is well described by a polynomial, but it will converge very slowly if this is not the case.

\subsection{Adaptive sampling}
When an integrand has structure on scales smaller than the integration interval, an adaptive integration scheme is often the best choice, since it will subdivide the interval until it resolves the structure and reach the required accuracy. We will use the 15 point Gauss-Kronrod quadrature formula as a basis for our adaptive integrator; 7 of the 15 points can be used to obtain a Gauss quadrature estimate of the integral, and the error estimate on the 15 point formula is then $\text{err}_\text{est.} = 200 |G7-K15|^{1.5}$~\cite{Kahaner:1989:NMS}.

The Gauss-Kronrod formula is defined on the open interval $(-1,1)$, but it can be rescaled to work on an arbitrary open interval $(a,b)$. We transform the indefinite integral into a definite integral by the substitution $x = \para{q+1}^{-1}$:
\begin{align}
\int_0^\infty \di q f\para{q} = -\int_0^1 \di x \frac{\di q}{\di x} f\para{q\para{x}} = \int_0^1 \di x x^{-2}f\para{q\para{x}}.
\end{align}
This integral can then be solved by the adaptive integrator. If the tolerance requirement is not met using the first 15 points, the interval is divided in two and the quadrature method is called recursively on each subinterval.

This method is very efficient when the integrand is smooth. For
practical purposes, this will be the case unless the phase-space
distribution is read from a file with sparse sampling: in this case,
the code must interpolate or extrapolate the file values in order to
cover the whole momentum range, and the next method may be more
efficient.

\subsection{Integration over tabulated distributions}

If some distribution function is not known analytically, but only on a
finitely sampled grid on $(\qmin,\qmax)$, we have to interpolate the
distribution function within the interval, and we have to extrapolate
the behaviour outside the interval. Inside the interval we use a
spline interpolation, while we assume $f(q<\qmin)\equiv f(\qmin)$
close to zero. For the tail, we assume the form $f(q) = \alpha
e^{-\beta q}$. Requiring the function and its first derivative to be
continuous at the point $q=\qmax$ leads to the following equations for
$\alpha$ and $\beta$:
\begin{align}
\alpha &= f(\qmax) e^{\beta \qmax}, \\
\beta &= -f(\qmax)^{-1} \left. \frac{\di f}{\di q}\right|_{q=\qmax}.
\end{align}
In the combined scheme we use the 4 point Gauss-Legendre method on the
interval $(0,\qmin)$, adaptive Gauss-Kronrod quadrature on
$(\qmin,\qmax)$ and the 6 point Gauss-Laguerre rule on the tail
$(\qmax,\infty)$\footnote{This version of the rule is obtained by a simple substitution.}. This scheme works well when the integrand is
interpolated from tabulated points.

\subsection{Implementation in CLASS}

When \CLASS{} initialises the background structure, it will find optimal momentum samplings for each of the species. More specifically, we start by computing the integral of the
distribution function multiplied by the test function at high accuracy, which gives a reference value which can be used for comparison. It also creates
a binary tree of refinements, from which we can extract integrals at various
levels, where level 1 is the best estimate. We choose the highest
possible level which results in an error which is less than the input
tolerance, and we extract the nodes and weights from that level. 

The code will now search for the lowest number of nodes required for computing the integral with the desired accuracy using Gauss-Laguerre quadrature. The most efficient method, the method using the lowest number of points, is then chosen. For a distribution not departing too much from a
Fermi-Dirac one, this will usually be Gauss-Laguerre quadrature.

The scheme suggested here has the benefit, that there is just one
tolerance parameter directly related to how well the integral is
approximated, \emph{independently} of the distribution
function. However, for this to be exactly true, we require the test
function to be a sufficiently realistic representation of $q^n \Psi_l$
for $n=2,3,4$ and $l=0,1,2$ for the perturbations\footnote{The energy $\epsilon$ behaves like $q$ in the relativistic limit and like a constant in the non-relativistic limit, so there is an intermediate range where it is not completely described by a polynomial of finite order. However, we do not think this error is a dominant one.}. We have tried different test functions, but in the end we found the polynomial
$t(q) = a_2q^2+a_3q^3+a_4q^4$ to be adequate. The coefficients were
chosen such that
\begin{align}
a_n \int_0^\infty \di q \frac{q^n}{e^q+1} &= 1.
\end{align}

When the phase-space distribution function is passed in the form of a
file with tabulated $(q_j, f_j)$ values, the code compares the three
previous methods (still with a common tolerance parameter) and keeps
the best one, which is usually the third one in the case of a poor
sampling of the function, or one of the other two in the opposite
case.

\section{\ncdm accuracy settings \label{accuracy}}
Taking the runs with accuracy settings {\tt
cl\_ref.pre} as a reference, we decrease the precision for each
parameter while keeping the $\Delta \chi^2$ roughly below a given
limit, chosen to be either 0.1 or 1. This exercise was already
performed in \cite{class_comp} for all parameters not related to
NCDM, leading to the definition of two precision files
{\tt chi2pl0.1.pre} and {\tt chi2pl1.pre} which are available on the
\CLASS{} web site. Here, we only need to set the NCDM precision
parameters in these two files to correct values. Our results are
listed in Table~\ref{settings}, in the second and third columns. They
take advantage of the fluid approximation, and use an extremely small
number of momenta (8 or 5 only). We checked that these settings
provide the correct order of magnitude for $\Delta \chi^2$ within a
wide range of neutrino masses, at least up to 2~\eV. This is shown in
Table~\ref{tab:perf} for the two cases {\tt chi2pl0.1.pre} and {\tt
chi2pl1.pre}, as well as for the case {\tt chi2pl1.pre} with the fluid
approximation removed.  Around $m=2$~\eV, the error induced by the
fluid approximation starts increasing significantly: when exploring
this region, the user should either turn off the approximation, or
increase the value of the $k\tau$ trigger. Given current limits on
active neutrino masses, the interesting mass range to explore is below
2~\eV, and in most projects, the \CLASS{} users can safely employ the default
settings of {\tt chi2pl0.1.pre} and {\tt chi2pl1.pre}
including the fluid approximation.

These settings are optimised for fitting the CMB spectra only. For the
matter power spectra, the files {\tt chi2pl0.1.pre} and {\tt
  chi2pl1.pre} produce an error of the order of a few per cents in the
range $k \in [0.05; 1]h\text{Mpc}^{-1}$ (for any neutrino mass and with/without
the fluid approximation). In order to get accurate matter power
spectra, it is better to employ the settings {\tt cl\_permille.pre},
{\tt cl\_2permille.pre}, {\tt cl\_3permille.pre}, which lead to a
precision of 0.1\%, 0.2\% or 0.3\% for $C_l^{TT}$ in the range
$2<l<3000$, even in the presence of neutrino masses. 
In these files, we fixed the fluid approximation trigger
to a rather larger value in order to get a precision of one
permille for the matter power spectrum for $k<0.2h\text{Mpc}^{-1}$ and
$m<2$~eV, or a bit worse for mildly non-linear scales $k\in
[0.2; 1]h\text{Mpc}^{-1}$. The power spectrum accuracy with such settings
is indicated in Table~\ref{pk_pre} for various values of the mass.

\TABLE{
\begin{tabular}{l|ccc}
 & {\tt cl\_ref.pre} & {\tt chi2pl0.1.pre} & {\tt chi2pl1.pre}\\
\hline
{\tt tol\_ncdm\_bg} & $10^{-10}$ & $10^{-5}$ & $10^{-5}$\\
{\tt tol\_ncdm} & $10^{-10}$ & $10^{-4}$ & $10^{-3}$\\
{\tt l\_max\_ncdm} & 51 & 16 & 12\\
fluid approximation & none & {\tt ncdmfa\_class} & {\tt ncdmfa\_class} \\
$k\tau$ trigger & -- & 30 & 16 \\ 
\hline
number of $q$ (back.) & 28 & 11 & 11\\
number of $q$ (pert.) & 28 & 8 & 5 \\
number of neutrino equations & 1428 & 136$\rightarrow$3 &  65$\rightarrow$3
\end{tabular}
\caption{Accuracy parameters related to NCDM in the three precision files {\tt cl\_ref.pre}, {\tt chi2pl0.1.pre} and {\tt chi2pl1.pre}. When the fluid approximation is used, the method described in section \ref{ncdmfa} is employed, and the switching time is set by the above values of $k\tau$. Below these parameters, we indicate the corresponding number of momenta sampled in background quantities and in perturbation quantities, as well as the number of neutrino perturbation equations integrated over time, equal to $({\tt l\_max\_ncdm}+1)$ times the number of sampled momenta when the fluid approximation is not used, and to three afterwards.
}
\label{settings}
}

\TABLE{
\begin{tabular}{c|ccc}
mass (\eV)& {\tt chi2pl0.1.pre} & {\tt chi2pl1.pre} & same without approx. \\
\hline
$10^{-3}$& 0.087 & 0.94 & 0.90\\
$10^{-2}$& 0.087 & 0.93 & 0.92\\
0.1  & 0.092 & 0.90 & 0.92\\
1    & 0.083 & 0.96 & 0.82\\
2    & 0.157 & 1.10 & 0.93\\
\end{tabular}
\caption{For a CMB instrument with the sensitivity of
Planck, $\chi^2$ difference between the spectra obtained with
reference accuracy settings and with degraded accuracy settings, for
various values of the neutrino mass (all models have two massless and
one massive neutrinos). This shows that our accuracy settings {\tt
chi2pl0.1.pre} and {\tt chi2pl1.pre} always lead to an accuracy of
roughly $\Delta \chi^2 \sim 0.1$ or $\Delta \chi^2 \sim 1$
respectively. The last column correspond to the settings of {\tt
chi2pl1.pre}, but without the fluid approximation.}
\label{tab:perf}
}
\TABLE{
\begin{tabular}{c|cc}
mass (\eV) & $k<0.2h\text{Mpc}^{-1}$
& $k\in [0.2; 1]h\text{Mpc}^{-1}$ \\
\hline
$10^{-3}$  & 0.04\% &  0.12\% \\
$10^{-2}$  & 0.04\% &  0.12\% \\
0.1  & 0.05\% &  0.12\% \\
1    & 0.06\% &  0.8\% \\
2    & 0.2\% &  1.5\% \\
\end{tabular}
\caption{Maximum error induced by any of the {\tt cl\_permille.pre},
  {\tt cl\_2permille.pre} or {\tt cl\_3permille.pre} precision
  settings on the linear matter power spectrum $P(k)$, for
  approximately linear scales $k<0.2h\text{Mpc}^{-1}$ (first column) or mildly
  non-linear scales $k \in [0.2; 1]h\text{Mpc}^{-1}$ (second column), and for
  various values of the neutrino mass (all models have two massless
  and one massive neutrinos). The fluid approximation introduces an
  error which remains below the 0.1\% level until $k=0.2h\text{Mpc}^{-1}$ for
  $m<2$~eV, and exceeds this level for larger masses.}
\label{pk_pre}
}

\bibliographystyle{utcaps}
\bibliography{NCDM}

\end{document}